\documentclass[twocolumn,aps,prc,eqsecnum,tightenlines,floats,floatfix,nofootinbib]{revtex4}

\usepackage{graphicx}

\def\bea{\begin{eqnarray}}
\def\eea{\end{eqnarray}}

\def\sfrac#1#2{{\textstyle \frac{#1}{#2}}}

\newcommand{\braket}[2]{\langle #1|#2\rangle}

\def\be{\begin{equation}}
\def\ee{\end{equation}}
\def\ba{\begin{eqnarray}}
\def\ea{\end{eqnarray}}

\usepackage{epstopdf}
\usepackage{graphics}
\usepackage{graphicx}
\usepackage{amsmath}
\usepackage{amssymb}

\begin{document} 

\phantom{0}
\vspace{-0.2in}
\hspace{5.5in}
\parbox{1.5in}{ } 

\vspace{-1in}

\title
{\bf Shape of the  $\Delta$ baryon in a covariant spectator quark model}
\author{G.~Ramalho,$^{1}$ 
M.~T.~Pe\~na,$^{1,2}$ and A.~Stadler$^{3,4}$ 
\vspace{-0.1in}  }

\affiliation{
$^1$CFTP,  Instituto Superior T\'ecnico, 
Universidade T\'ecnica de Lisboa,
Avenida Rovisco Pais, 
1049-001 Lisboa, Portugal \vspace{-0.15in}}
\affiliation{
$^2$Department of Physics, Instituto Superior T\'ecnico, 
Universidade T\'ecnica de Lisboa,
Av.\ Rovisco Pais, 
1049-001 Lisboa, Portugal \vspace{-0.15in} }
\affiliation{
$^3$Departamento de F\'isica, Universidade de \'Evora,
7000-671 \'Evora, Portugal\vspace{-0.15in} }
\affiliation{$^4$Centro de  F{\'\i}sica Nuclear da Universidade de Lisboa, 
  1649-003 Lisboa, Portugal}
   \vspace{0.2in}
\date{\today}

\phantom{0}

\begin{abstract}
Using a covariant spectator quark model that describes the
recent lattice QCD data for the $\Delta$ electromagnetic 
form factors and all available experimental data 
on $\gamma N \to \Delta$ transitions,  
we analyze the charge and magnetic dipole
distributions of the $\Delta$ baryon and discuss its shape.
We conclude that the quadrupole moment of the 
$\Delta$ is a good indicator of the deformation 
and that the $\Delta^+$ charge distribution 
has an oblate shape.
We also calculate transverse moments 
and find that they do not lead to unambiguous 
conclusions about the underlying shape. 
\end{abstract}

\vspace*{0.9in}  
\maketitle

\section{Introduction}

The determination of the shape of a baryon is an interesting problem 
that addresses
complex technical issues, both from the experimental and 
theoretical point of view \cite{Buchmann00,Alexandrou12}.
Recently, for example, there has been particular interest (and controversy) 
about possible deviations from a spherically symmetric shape 
of the nucleon. 
In this work we will address another interesting question, 
namely whether the $\Delta$ baryon 
has a spherical shape or not, and maybe more importantly, what we mean by that.

Information on a baryon's shape is encoded in its
electromagnetic form factors, which can be
measured, at least in principle. 
The electromagnetic form factors describe how a particle 
interacts with a photon, and the number of independent 
form factors depends on the particle's spin. 
A particle with spin-1/2, such as the nucleon, 
is characterized by only two form factors, namely the electric charge 
$G_{E0}$ and magnetic dipole $G_{M1}$, whereas
a particle with
spin-3/2, such as the $\Delta$, 
has altogether four form factors, namely, in addition 
to $G_{E0}$ and $G_{M1}$, also electric quadrupole $G_{E2}$
and magnetic octupole $G_{M3}$ form factors. 

Higher-order form factors and moments can provide 
a measure 
of the deformation of an extended particle. 
For instance,  nonvanishing values of the electric quadrupole moment  
and of the magnetic octupole moment, which are proportional to $G_{E2} (0)$ and $G_{M3} (0)$, respectively, indicate a deviation of the
charge and magnetic dipole distributions from the spherically
symmetric form \cite{Buchmann08}.

At this point, it is important to distinguish between ``spectroscopic'' 
and ``intrinsic'' moments, the former being the 
observable values of the corresponding electromagnetic form factor 
at zero momentum transfer, whereas the latter refer to quantities
calculated from charge or magnetic density distributions, 
which are not directly observable \cite{Buchmann00,Alexandrou12}.
For instance, nucleons or spin-1/2 nuclei
do not possess an electric quadrupole or magnetic octupole moment, 
and therefore they cannot be deformed 
if finite higher spectroscopic moments are used as criteria 
for deformation.
On the other hand, their
intrinsic density distribution may not be spherically symmetric, 
giving rise to nonvanishing intrinsic higher moments 
\cite{Nucleon,Buchmann00,Alexandrou12}.

The situation is somewhat simpler in the case of the $\Delta$, 
the lightest baryon with spin-3/2 and first candidate for
a nonvanishing electric quadrupole moment, because 
its intrinsic and spectroscopic quadrupole moments differ 
only by a constant factor  \cite{Alexandrou09,Buchmann00}.
Thus, there is a strong motivation to estimate the $\Delta$ 
electric quadrupole moment. 
Previous
studies on the 
deformation of the $\Delta$ can be found in 
Refs.~\cite{Alexandrou12,Alexandrou09,Becchi65,Glashow79,Hulthage80,Isgur82,Gross83,Viollie83,Clement84,AlexandrouDeform,Nicmorus10a,Alexandrou08b}.

It is only  in the 
nonrelativistic limit that
the spatial distributions of the charge and magnetic
densities are related to the electromagnetic form factors through 
a Fourier transform, such that information about 
the shape of the density distributions can be 
accessed through form factor data: in the Breit frame, 
the charge distribution 
is the Fourier transform of $G_{E0}$, 
and the magnetic dipole density is
the Fourier transform of $G_{M1}$~\cite{HalzenMartin,Kelly02}. 

However, this interpretation has its limitations. 
In a relativistic description, 
the spatial distribution of charge or magnetic densities depends 
on the reference frame.
When an absorbed photon imparts only a small momentum to the baryon, 
this frame dependence can be ignored, but
in the general relativistic case there is no direct relation 
between form factors and coordinate-space densities.   
It was to address this difficulty that alternative concepts 
to measure deformation were proposed, such as transverse densities and moments 
\cite{Alexandrou09,Lorce09,Transverse1,Transverse2,Miller08} and
spin-dependent density distributions \cite{Miller03}.

Transverse density distributions were  introduced recently within the context of a light-front description of the electromagnetic current.  In this formulation, the charge and magnetic density distribution of a particle is described as seen from a light-front moving towards the particle. The longitudinal direction is effectively eliminated by projecting the density onto the transverse plane, which is not subject to Lorentz contraction.

For spin polarizations in the transverse plane, radially asymmetric transverse density distributions can be obtained, which may then be analyzed in terms of multipoles. However, 
it is not clear how exactly they are connected to the intrinsic deformation of the particle's density distribution in its rest frame. For instance, just as in a number of previous model calculations, 
a recent lattice QCD study found a negative value 
of 
the electric quadrupole moment of the $\Delta^+$ baryon 
in the state with spin projection 3/2, implying 
an oblate
intrinsic deformation relative to the spin direction, 
whereas the corresponding transverse charge density field 
pattern shows a prolate shape 
\cite{Alexandrou09}. 

It has been suggested that information about deformation should 
be deduced by comparing the transverse moments to the 
so-called ``natural'' moments of structureless, elementary particles.
In this formulation, a  particle is \emph{defined} as elementary if 
its light-cone helicity  is nontrivially
conserved at tree level.
The argument given to support this assumption is that in a composite particle the elementary constituents can jump between orbital states and thereby change the particle's total helicity, whereas this is not possible if the particle has no constituents  \cite{Lorce09}.

For such a structureless particle with spin 3/2 and charge
$e_\Delta$ (specified in units of the proton charge),  
the natural moments---labeled here with the superscript $(nat)$---are \cite{Lorce09} 
\be
G^{(nat)}_{E2}(0)= - 3 e_\Delta, \hspace{.5cm}
G^{(nat)}_{M3}(0)= - e_\Delta,
\label{eqGnatural}
\ee
besides
\be
G^{(nat)}_{E0}(0)= e_\Delta, \hspace{.6cm}
G^{(nat)}_{M1}(0)= 3 e_\Delta.
\label{eqGnatural2}
\ee
In the example of the $\Delta^+$ with spin projection 3/2, 
a negative sign of $G_{E2}(0)$ is consistent with 
a prolate transverse charge density 
distribution if $G_{E2}(0) > G^{(nat)}_{E2}(0)$. 
In general, the transverse electric quadrupole moment 
is only a function of the anomalous electromagnetic moments \cite{Lorce09}, 
and a  nonzero value indicates a deviation from 
a pointlike structure
when viewed from the light-front.

The question remains what the transverse moments can tell us about the intrinsic density deformation in the rest frame.
In the following we present the intrinsic three-dimensional rest-frame densities obtained in covariant model calculations of the $\Delta$ baryon, and we will return to this question when we discuss our results.

In this work we used relativistic quark-diquark model wave functions of the  $\Delta$ baryon that were constructed within the 
covariant spectator theory~\cite{Gross} 
to calculate charge and magnetic density distributions, both in momentum and coordinate space. Thus we have direct information about their shape. In particular, we compared two models, one which includes only $S$-waves and is therefore spherically symmetric, and another that includes $D$-wave components that induce a small deformation. We determined then how this shape information manifests itself in the higher moments, as well as in the corresponding transverse moments. 
We want to emphasize that our moments are not calculated 
at tree level and contain anomalous contributions. 
Therefore they do not reduce to the natural values 
of Eqs.~(\ref{eqGnatural}) and Eqs.~(\ref{eqGnatural2}).

With a spherically symmetric spatial wave function, 
where the quark and diquark are in a relative $S$-wave \cite{DeltaFF}, we obtained
\ba
G_{E2}(0)= 0, \hspace{.5cm} G_{M3}(0)=0.
\label{eqSp01}
\ea
This result does not depend on any specific model parameters, but holds in general as long
as only $S$-waves are present.

Once higher orbital angular momentum components are included 
in the quark-diquark wave function,
nonvanishing values of $ G_{E2}(0)$ and $G_{M3}(0)$ are generated
\cite{DeltaDFF,DeltaDFF2}. 
This indicates that it is possible 
to relate deviations from spherical symmetry in the charge 
(dipole moment) distribution to the value
of the electric quadrupole (magnetic octupole) moment,
similar to what was found in the nonrelativistic limit. 
The calculated values of the transverse quadrupole and octupole moments on the other hand do not seem to 
give any clear indication on the deformation of the density distributions.

This paper is organized as follows: The covariant 
spectator quark model is introduced 
in Sec.~\ref{secSpecQuarkModel}. 
In Sec.~\ref{secFFvsHA}, we relate 
the electromagnetic form factors with 
the experimentally accessible polarized helicity amplitudes.  
In Sec.~\ref{secTrans} 
we discuss the deformation of the $\Delta$ as determined from transverse densities, and in Sec.~\ref{secClassic}
we present the results obtained from the usual three-dimensional densities.
In Sec.~\ref{secConclusions} we draw our conclusions.

\section{Spectator quark model}
\label{secSpecQuarkModel}

We apply a quark model obtained in the covariant spectator formalism 
\cite{Nucleon,Gross,FixedAxis,Octet,OctetFF,InMedium,Nucleon2,nucleonR,ExclusiveR,Omega,GE2Omega,spin32}, 
and parameterized to describe
the $\Delta$ baryon,
as discussed in detail in
Refs.~\cite{NDelta,NDeltaD,LatticeD,Lattice,DeltaDFF,DeltaDFF2}.
The $\Delta$ wave function is
a mixture of an $S$ state ($L=0$) and two $D$ states 
($L=2$, coupled to core spin $1/2$ and $3/2$)
for the quark-diquark system \cite{NDeltaD,LatticeD}, of the general form 
\be
\Psi_\Delta=
N \left[
\Psi_S + a \Psi_{D3} + b \Psi_{D1} \right].
\label{eqPsiDel}
\ee
In this equation, $a$ is the admixture coefficient of the $D3$ state 
(quark core with spin-3/2) 
and $b$ the admixture coefficient of the $D1$ state
(quark core with spin-1/2).
The momentum and spin indices are suppressed for simplicity.
The $S$- and $D$-state wave function components are 
written in terms of spin, orbital angular momentum, and 
isospin operators, multiplied by 
scalar functions, $\psi_S$, $\psi_{D3}$, and 
$\psi_{D1}$. In our covariant spectator model, 
the diquark four-momentum is on-mass-shell, and therefore these scalar 
functions depend only on the square of the quark  
four-momentum, $(P-k)^2$, where 
$P$ and $k$ denote the
$\Delta$ and the diquark four-momentum, respectively.

Assuming each of the
$\Delta$ wave function components in (\ref{eqPsiDel}) to be normalized to 1, 
the overall normalization constant becomes  $N=1/\sqrt{1+ a^2+ b^2}$.
This specific form of the wave function
was introduced in Refs.~\cite{NDeltaD,LatticeD}
and two different parameterizations were studied  
in Refs.~\cite{DeltaDFF,DeltaDFF2}. 
Here we will use the model of Ref.~\cite{LatticeD},
because it gives a  
more consistent description of the valence quark 
contribution of the $\Delta$ to the $\gamma N \to \Delta$ reaction, 
both in the physical region and in the regimes accessible in
lattice QCD calculations with heavy pions.
In this model, the two $D$-state probabilities are both about 0.89\% 
($a=0.08556$ and $b=0.08572$). 
For more details on the model we refer to Refs.~\cite{LatticeD,DeltaDFF2}.

The internal structure of the constituent quarks 
is described in terms of quark electromagnetic form factors, 
parametrized through a vector meson dominance 
mechanism and included in an effective 
quark current $j_q^\mu$ \cite{Nucleon,Lattice,LatticeD,Omega}.
Employing the wave function (\ref{eqPsiDel}) 
and the quark current $j_q^\mu$,  
in the covariant spectator formalism 
we write the electromagnetic current~\cite{Nucleon,NDelta,DeltaFF} as
\ba
J^\mu &=& 
 3 \sum_{\lambda_s} \int_k 
\overline \Psi_\Delta (P_+,k;s^\prime) j_q^\mu \Psi_\Delta(P_-,k;s) 
 \label{eqCurrent1}  \\
  &=& 
-\bar u_\alpha (P_+,s^\prime) \left\{ \left[
F_1^\ast (Q^2) g^{\alpha \beta} + F_3^\ast (Q^2) 
\frac{q^\alpha q^\beta}{4M_\Delta^2} \right] \gamma^\mu 
\right. \nonumber \\
& & 
\left. 
+
\left[
F_2^\ast (Q^2) g^{\alpha \beta} + F_4^\ast (Q^2) 
\frac{q^\alpha q^\beta}{4M_\Delta^2} \right] 
\frac{i \sigma^{\mu \nu} q_\nu}{2 M_\Delta} \right\}u_\beta(P_-,s),
\nonumber 
\ea
where $P_+$ ($P_-$) represents 
the final (initial) four-momentum, 
$q=P_+ - P_-$ is the transferred momentum, $Q^2=-q^2$, $M_\Delta$ is the mass of the $\Delta$,
and 
${\lambda_s}$ the diquark polarizations. For the covariant integration over the on-mass-shell diquark momentum $k$ 
we use the abbreviation
\be
 \int_k \equiv \int \frac{d^3 k}{(2\pi)^32E_s} \, ,
\ee
with $E_s=\sqrt{m_s^2+{\bf k}^2}$, where $m_s$ is a model parameter 
that corresponds to a mean value of the spectator diquark mass 
\cite{Nucleon,Nucleon2}.
The asymptotic states $u_\alpha$ 
are the Rarita-Schwinger vector states \cite{Rarita41}. 
Throughout this paper we follow the convention used in 
our previous work that the diquark polarization indices, 
$\lambda_s$,  on the wave functions are suppressed.

The multipole $\Delta$ form factors can be written 
as linear combinations of $F_i^\ast$, $i=1,\dots,4$
\cite{DeltaFF,DeltaDFF2,Pascalutsa07,Nozawa90}.
For 
$Q^2=0$, to first order in the admixture coefficients $a$ and $b$, 
one finds \cite{DeltaDFF2} 
\ba
G_{E0}(0)  &=& N^2 e_\Delta    \nonumber\\
G_{M1}(0) &=&  
N^2 \left(e_\Delta+ \kappa_\Delta \right)           
\nonumber\\
G_{E2}(0) &=&  3 (aN^2) e_\Delta {\cal I}_{D3}^\prime
\nonumber\\
G_{M3}(0) &=& 
 \left(e_\Delta+ \kappa_\Delta \right)  
N^2  \left[ a\, {\cal I}_{D3}^\prime + 
2\, b \, {\cal I}_{D1}^\prime \right],
\label{eqDff}
\ea
where
\bea
e_\Delta= \sfrac{1}{2}(1+ \bar T_3),& 
\hspace{.7cm}
&\kappa_\Delta= \sfrac{1}{2}(\kappa_+ + \kappa_- \bar T_3)
\sfrac{M_\Delta}{M_N},
\nonumber\\
\kappa_+=2\kappa_u-\kappa_d,\;&  &\kappa_-=\sfrac23\kappa_u+\sfrac13\kappa_d,
\label{eqDefK}
\eea
with  $\bar T_3 = \mbox{diag}(3,1,-1,-3)$, 
and $M_N$ is the nucleon mass. 
The factors ${\cal I}_{D1}^\prime$ and ${\cal I}_{D3}^\prime$
are defined in terms of the overlap integrals
between the initial $S$-state and the final $D$-state, as
\ba
{\cal I}_{D3}^\prime
&=& \lim_{\tau \to 0} \frac{1}{\tau}
\int_k b(k,q,P_+)\psi_{D3} (P_+,k) \psi_S(P_-,k) 
\nonumber \\
{\cal I}_{D1}^\prime
&=& \lim_{\tau \to 0} \frac{1}{\tau}
\int_k  b(k,q,P_+) \psi_{D1} (P_+,k) \psi_S(P_-,k),
\nonumber 
\ea
with $\tau=\sfrac{Q^2}{4M_\Delta^2}$. The function  
$b(k,q,P_+)$, whose detailed form is given in Ref.~\cite{NDeltaD}, reduces to 
${\bf k}^2Y_{20}(\hat {\bf k})$ in the limit $Q^2 \to 0$,
where $Y_{20}(z)$ is the familiar spherical harmonic.   

The model described above was applied in Ref.~\cite{DeltaDFF2} 
to calculate the $\Delta$ electromagnetic 
form factors, and its results were compared successfully to the recent 
lattice QCD simulations 
of Refs.~\cite{Alexandrou09,Boinepalli09}. 
Although the model is still incomplete 
because important degrees of freedom, such as
meson (pion in particular) cloud effects, 
are not included, it
agrees well with the 
lattice QCD data 
for $G_{E0}$ and $G_{M1}$ \cite{Alexandrou09,Boinepalli09}
and is also consistent with the unquenched $G_{E2}$ 
data \cite{Alexandrou09}. 
This success 
can be due to
an effective suppression of pion cloud effects 
in the 
 $\gamma \Delta \to \Delta$ reaction\footnote{
A similar effect can be seen in 
the nucleon form factors.
In some models, the pion cloud contributions 
are around 10\% \cite{OctetFF,InMedium}.
Even in models where pion cloud contributions 
to the nucleon magnetic moments 
are significant ($\approx 40\%$)  \cite{Octet}, 
the difference between the results with the pion cloud and the results when the pion cloud effect is removed differ by  only
about 5\%  \cite{Octet,OctetFF}.},
in contrast to 
the  $\gamma N \to \Delta$ transition 
where the opening of the $\pi N$ channel is crucial
\cite{NDelta,NDeltaD,LatticeD}.
It is also possible that effective pion cloud effects 
are already included adequately through the vector meson dominance mechanism
which models the effective quark current $j_q^\mu$.

\section{Form factors and helicity amplitudes}
\label{secFFvsHA} 

The electromagnetic form factors of a baryon
are invariant 
functions of $Q^2$. They are independent of the 
reference frame and of the initial 
or final polarization of the baryon. 
Note that these functions are not directly measured in an experiment. 
What can be measured are cross sections in a particular frame, from which 
helicity transition 
amplitudes between two different or equal 
polarization states of the baryon can be deduced.

In a process like $\gamma \Delta \to \Delta$, 
there are only 3 independent components of the current, 
as a consequence of current conservation.
These components can be chosen to be $J^0,J^x$ and $J^y$, 
or, alternatively, $J^0,J^+$ and $J^-$, 
with $J^\pm \equiv \mp \frac{1}{\sqrt{2}}(J^x\pm i J^y)$.
Note that $J^\pm$ is associated with the 
photon polarizations $\lambda=\pm$ that 
involve a change of the baryon polarization ($\pm 1$),
and $J^0$ with $\lambda=0$  where the 
baryon polarization is conserved.

The transition amplitude
for spin projections $s$ and $s^\prime$
is
\be
J^{\lambda}(s^\prime,s) 
= -
\bar u_\alpha (P_+,s^\prime) \left[
{\cal O}^{\alpha \beta \mu} 
(\epsilon_\mu)^{\lambda} \right]
u_\beta(P_-,s),
\ee
where the operator ${\cal O}^{\alpha \beta \mu}$
is implicitly defined through Eq.~(\ref{eqCurrent1}), 
and $(\epsilon_\mu)^{\lambda}$ are the photon
polarization vectors, with
$(\epsilon_\mu)^0= (1,0,0,0)$ and  
$(\epsilon_\mu)^\pm = \pm \frac{1}{\sqrt{2}}(0,1,\pm i,0)$.

We will work in the Breit frame, where
the photon four-momentum is 
$q=(0,0,0,Q)$, with $Q=\sqrt{Q^2}$, 
the photon  three-momentum 
${\bf q}$ points along the positive $z$-direction, 
and the initial and final total momenta are $P_\pm=\left(M_\Delta \sqrt{1+ \tau},0,0,\pm \sfrac{1}{2} Q \right)$. 

The spin nonflip components 
of the current ($s^\prime=s$)
are
\cite{Alexandrou09}\footnote
{In Refs.~\cite{Alexandrou09} the normalization is
  $\bar u_\alpha(P,s) u^\alpha(P,s) = - 2M_\Delta$ 
  for $P=(M_\Delta, {\bf 0})$.
  Here we use $\bar u_\alpha(P,s) u^\alpha(P,s) = - 1$.}
\be
J^0\left(s,s \right) =
G_{E0}(Q^2) - \frac{2}{3} f_s(s) \tau G_{E2}(Q^2), 
\label{eqJ0BF}
\ee
for $s=\pm \sfrac{1}{2}, \pm \sfrac{3}{2}$, with 
$f_s(\pm \sfrac{3}{2}) \equiv 1$ and $f_s(\pm \sfrac{1}{2}) \equiv -1$.

One can combine the two independent amplitudes 
as a symmetric combination 
of matrix elements
\be
J^0_S = \frac{1}{2} 
\left[J^0\left( +\sfrac{3}{2}, +\sfrac{3}{2}\right) +
J^0\left( +\sfrac{1}{2}, +\sfrac{1}{2}\right) 
\right]= G_{E0}(Q^2),
\ee
and  an asymmetric combination
\be
J^0_A = \frac{1}{2} 
\left[J^0\left(+ \sfrac{3}{2}, +\sfrac{3}{2}\right) -
J^0\left( +\sfrac{1}{2}, +\sfrac{1}{2}\right) 
\right]= - \frac{2}{3} \tau G_{E2}(Q^2).
\ee
In the limit $Q^2 \to 0$, 
the first equation yields $G_{E0}(0)$, whereas 
$G_{E2}(0)$ cannot be obtained directly from 
the amplitudes at $Q^2=0$ because $\tau$ goes to zero.

Similarly, the magnetic form factors are obtained from the spin-flip current matrix elements
$J^\pm(s^\prime,s)$.
Again, there are only two independent amplitudes 
related with $G_{M1}$ and $G_{M3}$ for $Q^2 \ne 0$ 
(see Refs.~\cite{Alexandrou09} for details).

In a nonrelativistic formalism, the baryon's shape, 
and in particular any possible deformation---deviation 
from a spherically symmetric form---would depend
on the spin projection along the $z$-axis.
One can then define an electric 
charge distribution, $\rho_E({\bf r},s)$, associated with each 
spin projection ($\pm \sfrac32$ or $\pm \sfrac12$)
and define an intrinsic electric quadrupole momentum  
\cite{Buchmann00} as 
\be
Q_\Delta(s)= \int d^3 {\bf r} \, \rho_E({\bf r},s) \,
{\bf r}^2 Y_{20}(\hat {\bf r}) \, .
\ee
When $Q_\Delta(s) \neq 0$, its sign indicates whether the system 
is oblate ($Q_\Delta(s) < 0$) or prolate ($Q_\Delta(s) > 0$)
if $e_\Delta >0$ 
(the opposite shape if $e_\Delta <0$).
Note in particular that the shapes for $s=+ \sfrac32$ and for $s=+ \sfrac12$ can be different 
(see for instance Ref.~\cite{Gross83}).
Whenever the quadrupole magnetic moment 
is referred to without explicitly mentioning the polarization state, 
the maximum projection is assumed
\cite{Buchmann00}.

\section{Transverse density deformation}
\label{secTrans}

The interpretation of electromagnetic form factors as Fourier transforms of charge and magnetic distribution densities is valid only in the nonrelativistic limit. If the absorbed photon imparts a significant momentum transfer to the struck system,  the boost of the final state wave function relative to the one of the initial state can no longer be neglected, which spoils this simple interpretation. In order to still be able to extract information about distribution densities from the measured form factors, the concept of a transverse density distribution was introduced
\cite{Transverse1,Transverse2,Miller08}.

By going to an infinite momentum frame, 
the dependence on the longitudinal 
component of the momentum is eliminated, and the deformation
is defined in terms of densities 
in the space of the two transverse 
impact parameters, $b_x$ and $b_y$.
Taking the transverse spin projection 
($s_\perp = \pm \sfrac12, \pm \sfrac32$) oriented in the $x$ direction, 
the transverse electric quadrupole moment
${\cal Q}_\Delta^\perp(s_\perp)$
is \cite{Alexandrou09}
\ba
{\cal Q}_\Delta^\perp(s_\perp=+\sfrac{3}{2}) &=&
\frac{1}{2} \left\{ 
2 \left[ G_{M1}(0) - 3 e_\Delta 
\right]+ \frac{}{}
\right. \nonumber \\
&  &
\left. 
\left[
G_{E2}(0)+ 3 e_\Delta
\right] \frac{}{} \hspace{-.1cm}
\right\}\left(\frac{e}{M_\Delta^2}\right).
\label{eqQt}
\ea
For $s_\perp =+ \sfrac12$ one obtains
${\cal Q}_\Delta^\perp(+\sfrac{1}{2})= - {\cal Q}_\Delta^\perp(+\sfrac{3}{2})$.

In the previous equation, $[G_{M1}(0) - 3 e_\Delta]$
is interpreted as the electric quadrupole moment induced by the magnetic moment in the light-front frame,
and $[G_{E2}(0)+ 3 e_\Delta]$ 
as the polarization  effect due to the internal structure, also seen in that frame \cite{Alexandrou09}.
Using the natural values for $G_{E2}(0)$ and $G_{M1}(0)$,
Eqs.~(\ref{eqGnatural})--(\ref{eqGnatural2}),
from Eq.~(\ref{eqQt}) we conclude 
that ${\cal Q}_\Delta^\perp (+\sfrac{3}{2})=0$ 
for a structureless particle with spin 3/2 
seen from the light-front.

\begin{table}
\begin{center}
\begin{tabular}{l c c c c }
\hline
\hline
       & &  ${\cal Q}_\Delta^\perp\left(+\sfrac{3}{2}\right)$ & & 
${\cal O}_\Delta^\perp\left(+\sfrac{3}{2}\right)$     \\
\hline
{\bf 
Lattice QCD:} & &  & & \\
Quenched \cite{Alexandrou09} && 0.83$\pm$0.21 && \\
Wilson \cite{Alexandrou09} 
            & & 0.46$\pm$0.35 && \\
Hybrid \cite{Alexandrou09} 
            & & 0.74$\pm$0.68 && \\
{\bf Spectator quark models:} & &  & & \\
Spectator-S
\cite{DeltaFF} & &   0.29    & & -3.44   \\
Spectator-SD \cite{DeltaDFF2} & &   0.92    & & -3.38  \\ 
\hline
\hline
\end{tabular}
\end{center}
\caption{Transverse electric quadrupole moment 
${\cal Q}_\Delta^\perp\left(+\sfrac{3}{2}\right)$ in 
units of $\sfrac{e}{M_\Delta^2}$, and transverse 
magnetic octupole moment ${\cal O}_\Delta^\perp\left(+\sfrac{3}{2}\right)$ in 
units of $\sfrac{e}{2 M_\Delta^3}$, 
for the $\Delta^+$.}
\label{tableQdel}
\end{table}

Similarly, the transverse magnetic octupole moment becomes
\cite{Alexandrou09}
\ba
{\cal O}_\Delta^\perp (s_\perp = + \sfrac{3}{2}) &=&
\frac{3}{2} \left\{ 
- G_{M1}(0)  -G_{E2}(0) +
\frac{}{}
 \right. \nonumber \\
&  & 
\left.
G_{M3}(0) + e_\Delta 
\frac{}{} \hspace{-.1cm}
\right\}
\left(\frac{e}{2 M_\Delta^3}\right).
\label{eqOt}
\ea
For $s_\perp= + \sfrac12$, one has 
${\cal O}_\Delta^\perp(+\sfrac{1}{2})= 
- 3 {\cal O}_\Delta^\perp(+\sfrac{3}{2})$.
A spin-3/2 particle without structure in the light-front
has a vanishing 
transverse magnetic octupole moment.

The authors of Refs.~\cite{Alexandrou09}
suggest that the electric quadrupole 
and magnetic octupole moments of the $\Delta$ should 
be compared with their {\it natural} values, 
given in Eqs.~(\ref{eqGnatural})--(\ref{eqGnatural2}), 
and that deformation should be defined in terms of a 
positive or negative deviation from those reference 
values. Equations~(\ref{eqQt})--(\ref{eqOt})
depend indeed on these differences, but what they describe is not the full
deformation in three-dimensional coordinate space. 
Instead, because the transverse density is defined in
the two-dimensional $(b_x,b_y)$ plane, they 
measure an asymmetry of the density between the spin direction (along the $x$-axis)
and the perpendicular direction (along the $y$-axis) in the $xy$-plane.

To see how much information about the deformation of the 
$\Delta$ is contained in 
these higher transverse moments, we have calculated them for 
two covariant spectator quark-diquark models with significantly different shapes. 
The first, model II of Ref.~\cite{NDelta}, includes only $S$-states in the 
quark-diquark wave function, which is therefore spherically symmetric.
We call it here model ``Spectator-S''.
The second model, presented in Ref.~\cite{LatticeD}, 
is deformed, because apart from $S$-states it
includes also $D$-states. We refer to it here as model ``Spectator-SD''.

The spherical model  
Spectator-S yields \cite{DeltaFF} for the $\Delta^+$
\be
G_{E0}(0)=1, \hspace{.5cm} G_{M1}(0)=3.29,
\ee
and the quadrupole and octupole moments vanish.
For model Spectator-SD  one obtains
\cite{DeltaDFF2}
\ba
& &
G_{E0}(0) \simeq 1, \hspace{1.2cm} G_{M1}(0)= 3.27 \nonumber \\
& &
G_{E2}(0) =-1.70, \hspace{.5cm} G_{M3}(0)= -1.72. 
\label{eqvalues}
\ea  

From these values we can calculate the transverse 
electric quadrupole and the magnetic octupole moments.
The results are presented in Table \ref{tableQdel}, together with
lattice QCD data obtained by the 
MIT-Nicosia group \cite{Alexandrou09}
with three different methods, for pion masses in the range
$m_\pi =350-410$ MeV.
The positive sign of 
${\cal Q}_\Delta^\perp\left(+\sfrac{3}{2}\right)$
for all lattice calculations 
suggests a transverse distribution elongated in the 
spin direction. 
The same transverse deformation is produced by model Spectator-SD.
However, the transverse quadrupole moment for the pure $S$-wave model 
Spectator-S is not zero as one might expect.
Instead, it also predicts 
a deformation in the spin direction 
although not so strong as in the previous case.
Thus, whereas zero or nonzero values of electric quadrupole 
and magnetic octupole moments distinguish 
clearly between spherical and deformed $\Delta$ states, 
the corresponding transverse moments 
do not provide the same information. 
This is a quantitative illustration that the transverse moments 
are not an unambiguous  measure of deformation.

As for the transverse octupole moment, our result, 
${\cal O}_\Delta^\perp \left(+\sfrac{3}{2}\right) < 0$, 
suggests a deformation
perpendicular to the spin axis. 
This is true for both Spectator quark models, with and without 
$D$-states. 
The numerical values are very close, which means that  
${\cal O}_\Delta^\perp$ 
does not discriminate much between models with spherical or deformed 
wave functions.

At the moment, no lattice calculations of
${\cal O}_\Delta^\perp$ are available. 
But we can use the form factor data of   
the Adelaide group \cite{Boinepalli09}
at $Q^2=0.23$ GeV$^2$ in
Eq.~(\ref{eqOt}), replacing  $e_\Delta$ by 
$G_{E0}(Q^2)$, for a rough estimate.  We obtain
${\cal O}_\Delta^\perp(+\sfrac{3}{2})= 
\left( -23.8 \pm 22.3\right) \sfrac{e}{2M_\Delta^3}$,
which is consistent 
with our result---although with large statistical uncertainty---
and also suggests a deformation perpendicular to the spin direction.
For a more rigorous comparison with our predictions we have 
to wait for future lattice calculations.

\section{Deformation in spectator quark models}
\label{secClassic}

In this section, we calculate the $\Delta$ charge densities from wave functions obtained in a covariant spectator quark model, and we illustrate to what extent the distortion caused by the $D$-wave contributions manifests itself both in the momentum-space and coordinate-space densities. 
There is no need to make use of electromagnetic form factors to characterize deformation in this case, because the densities are calculated directly from the wave functions. Thereby we sidestep the usual problems 
in relating densities and form factors, namely that the latter involve wave functions in different reference frames. The former is the Fourier transform of the latter only in the nonrelativistic limit, whereas we are interested in the general, relativistic case. 

Because the covariant spectator theory is more naturally formulated in momentum space, in the following we start our discussion of $\Delta$ charge densities in the momentum-space representation where the wave functions 
were developed~\cite{Nucleon,NDeltaD}. Then we perform a Fourier transform of the wave function and discuss the densities in coordinate space.

\begin{figure*}[t]
\centerline{
\mbox{
\includegraphics[width=2.0in]{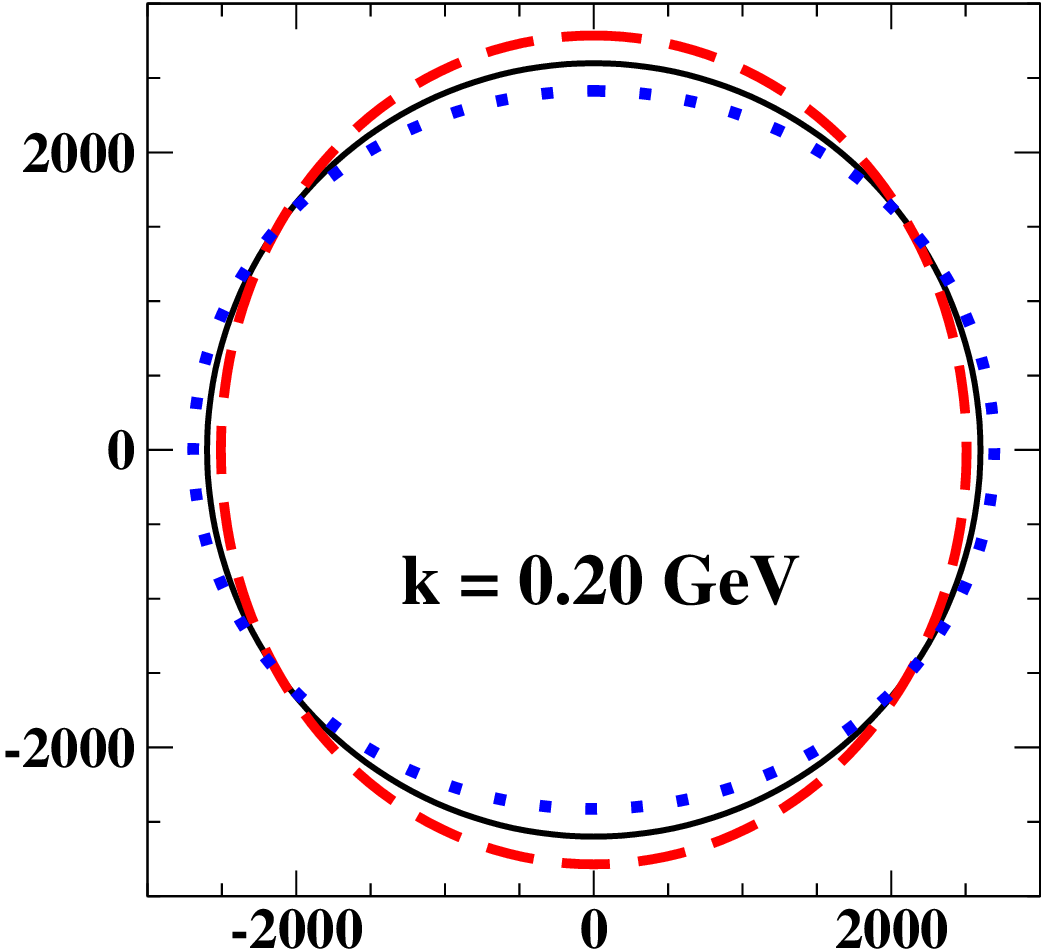} \hspace{1cm}
\includegraphics[width=1.85in]{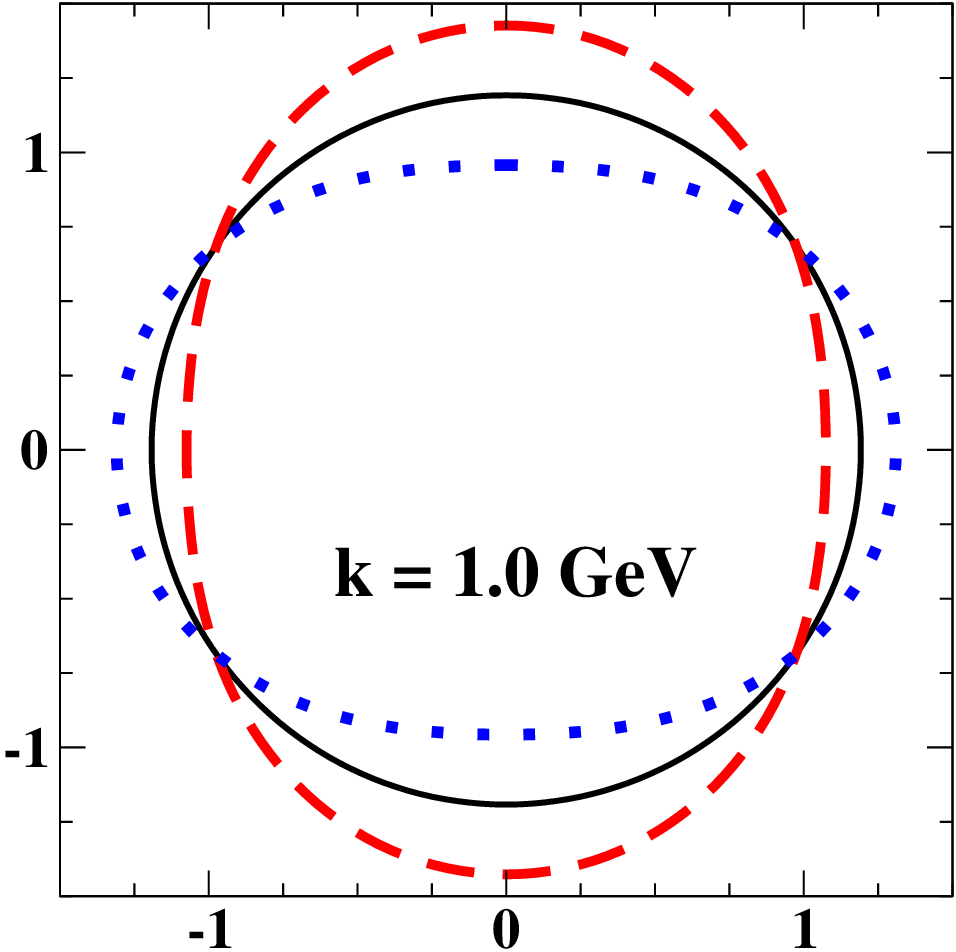} \hspace{1cm}
\includegraphics[width=2.0in]{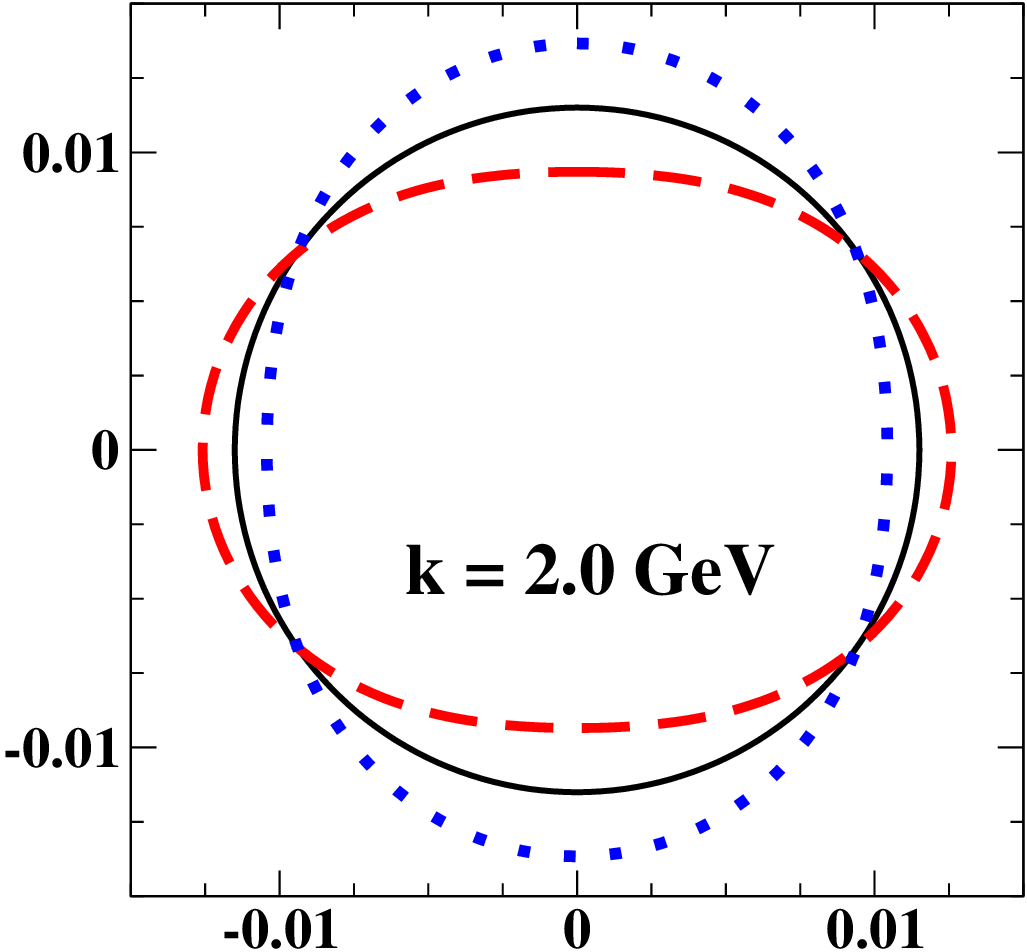}}}
\caption{\footnotesize{
Polar plots of $\rho_\Delta({\bf k},s)$ for three fixed values 
of ${\mathrm k}=|{\bf k}|$. In each case, 
the solid line represents $\rho_\Delta^S({\bf k})$,
the dashed line  $\rho_\Delta\left({\bf k},+\sfrac32\right)$, 
and the dotted line $\rho_\Delta\left({\bf k},+\sfrac12 \right)$. The scale for
$\rho_\Delta({\bf k},s)$ along the $k_x$ and $k_z$ axes is in units of GeV$^{-2}$. }}
\label{figKall}
\end{figure*}

\subsection{Momentum space}

The components of the 
$\Delta$ wave function of Eq.~(\ref{eqPsiDel})
can be written \cite{NDeltaD} as
\ba
& &
\Psi_S(P,k;s)= -\psi_S(P,k) (\varepsilon_P^\ast)^\alpha \; u_\alpha (P,s) \, ,
\label{eqPsiS}\\
& &
\Psi_{D3}(P,k;s)= \psi_{D3}(P,k) \Phi_{D3}(P,k;s) \, ,\\ 
& &
\Psi_{D1}(P,k;s)= \psi_{D1}(P,k) \Phi_{D1}(P,k;s) \, ,
\label{eqPsiD1}
\ea 
where the isospin state, which is a common factor in all wave functions, is omitted, and we use the notation
\be
\Phi_{D(2S)}(P,k;s)=
- 3 (\varepsilon_P^\ast)^\alpha 
\left({\cal P}_{S}\right)_{\alpha \beta} {\cal D}^{\beta \sigma} u_\sigma(P,s),
\label{eqPsiDX}
\ee
where $S=\sfrac12,\sfrac32$ is the core spin and $P$ the four-momentum of the $\Delta$.
In Eqs.~(\ref{eqPsiS}) and (\ref{eqPsiDX}), $\varepsilon_P^\ast$ 
is the diquark polarization vector in the 
fixed-axis representation~\cite{FixedAxis}. 
In the last equation, ${\cal P}_S$ is a 
projector onto the state $S= \sfrac12$ or $S= \sfrac32$, 
and ${\cal D}$ is the spectator $D$-state operator.
More details can be found in Ref.~\cite{NDeltaD}.

The scalar functions $\psi_S$, $\psi_{D3}$, and $\psi_{D1}$ regulate 
the momentum distribution 
of the quark-diquark system.
Because both the $\Delta$ baryon and its diquark constituent are 
on-mass-shell, these functions depend only on \mbox{$(P-k)^2$}. 
It is convenient to express them in 
terms of the dimensionless variable 
\ba
\chi= \frac{(M_\Delta-m_s)^2-(P-k)^2}{M_\Delta m_s},
\ea
where $m_s$ is the diquark mass.
Following Ref.~\cite{LatticeD}, 
we use the parametrizations
\ba
& &
\psi_S(P,k)=\frac{N_S}{m_s(\alpha_1+ \chi)^3} \\
& &
\psi_{D3}(P,k)=\frac{N_{D3}}{m_s^3(\alpha_2+ \chi)^4} 
\label{eqPsiRD3}
\\
& &
\psi_{D1}(P,k)=\frac{N_{D1}}{m_s^3}
\left\{\frac{1}{(\alpha_3+ \chi)^4} 
-\frac{\lambda_{D1}}{(\alpha_4+ \chi)^4}  \right\}.
\label{eqPsiRD1}
\ea
The momentum range parameters $\alpha_i$ ($i=1,\dots,4$), given in units of $m_s$, 
determine the long- and short-range dependence 
of the wave functions in coordinate space.
The normalization constants  
$N_S$, $N_{D3}$, and $N_{D1}$ are 
determined by the conditions 
$\int_k |\psi_S(\bar P,k)|^2=1$, 
$\int_k {\bf k}^4|\psi_{D3}(\bar P,k)|^2=1$, and
$\int_k {\bf k}^4|\psi_{D1}(\bar P,k)|^2=1$,
where $\bar P=(M_\Delta,{\bf 0})$ is the total four-momentum of the $\Delta$ baryon in its rest frame.
The coefficient $\lambda_{D1}$ is determined through 
the orthogonality between the $\Delta$ 
and nucleon states \cite{Nucleon,NDeltaD,LatticeD}.

In this work, we use the model of Ref.~\cite{LatticeD}, with the parameters
$\alpha_1=0.33660$, $\alpha_2=0.35054$, 
$\alpha_3= 0.33773$, and $\alpha_4=0.34217$.
One obtains $\lambda_{D1}=1.031898$, and the $D$-state admixture coefficients are
$a=0.08556$ and  $b=0.08572$. 

Similarly to the case of the nucleon in Ref.~\cite{Nucleon}, the momentum-space charge density
of the $\Delta$ in its rest frame is defined as
\be
\rho_\Delta({\bf k},s)=
\sum_{{\lambda_s}}
\Psi_\Delta^\dagger (\bar P,k;s) 
j_q
\Psi_\Delta (\bar P,k;s) \, ,
\label{eqRho}
\ee
where  
$s= \pm \sfrac12,\pm \sfrac32$ is the spin 
projection of the $\Delta$ state, 
and $j_q= 3j_1=\sfrac{1}{2} + \sfrac{3}{2} \tau_3$
the charge operator
\cite{Nucleon,NDelta}. Remember that implicitly the wave function $\Psi_\Delta (\bar P,k;s)$ depends also on the diquark polarization $\lambda_s$.

\begin{figure}[t]
\vspace{.3cm}
\centerline{
\mbox{
\includegraphics[width=3.3in]{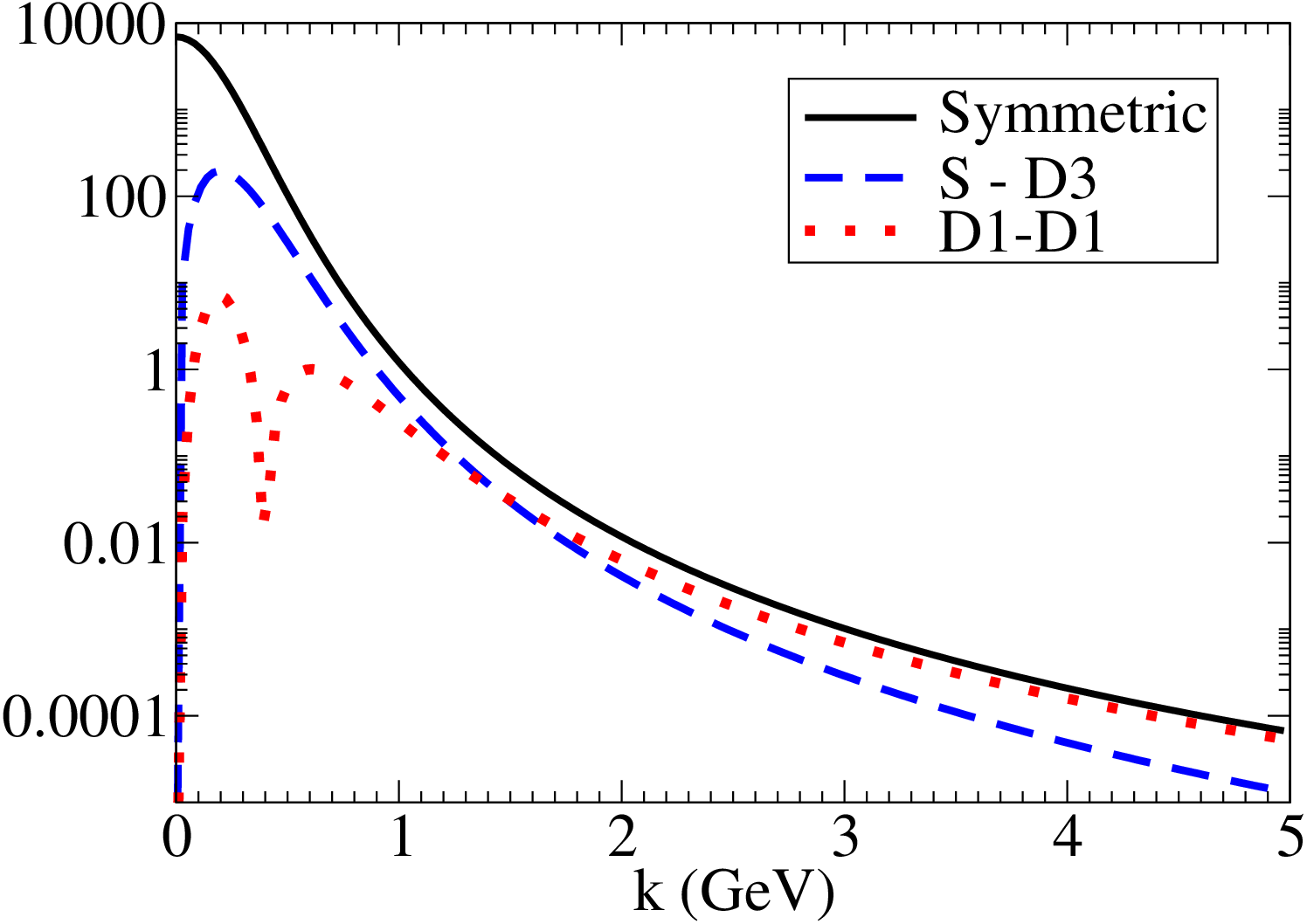}}}
\caption{\footnotesize{
Comparison of the three contributions to the total momentum-space density $\rho_\Delta({\bf k},s)$  in Eq.~(\ref{eqRhoTotal}) 
in units of GeV$^{-2}$.
The solid line represents the symmetric contribution, 
$\rho^S_\Delta ({\bf k})$, the dashed and dotted lines show 
the coefficients of $\hat{Y}_{20}(z)$ 
proportional to $\psi_S(\bar P,k) \psi_{D3}(\bar P,k)$ and $|\psi_{D1}(\bar P,k)|^2$, respectively. 
In all cases,
 the common factor $e_\Delta=1$, 
and only the absolute values are plotted.
}}
\label{figRho}
\end{figure}

Substituting (\ref{eqPsiDel}) into (\ref{eqRho})  
one gets 
\ba
& &
\rho_\Delta({\bf k},s)=
N^2 \rho_{\Delta, S}({\bf k},s) 
\nonumber \\
& &
\qquad
+  a^2 N^2 \rho_{\Delta ,D3}({\bf k},s)+ 
b^2 N^2 \rho_{\Delta ,D1}({\bf k},s)   \nonumber \\
& & 
\qquad
+ 2 aN^2 \rho_{\Delta ,SD3} ({\bf k},s),
\label{eqRho1}
\ea
where 
\be
\rho_{\Delta ,X}({\bf k},s)=
\sum_{\lambda_s} \Psi_X^\dagger (\bar P,k; s) j_q \Psi_X( \bar P,k; s),
\label{eqRhoX}
\ee
for $X=S,D3,D1$, and
\be
\rho_{\Delta ,S D3}({\bf k},s)=
\sum_{\lambda_s} \Psi_{D3}^\dagger (\bar P,k; s) j_q \Psi_S( \bar P,k ; s).
\label{eqRhoSD3}
\ee
The $S$ and $D3$ states can either be in the initial or 
final state, hence the factor of 2 in 
the last term of Eq.~(\ref{eqRho1}).

After performing the spin and isospin algebra, it is helpful to isolate the spherically symmetric contribution $\rho_\Delta^S$ from the angle-dependent terms in the total density, and one obtains
\ba
& &
\rho_\Delta({\bf k},s) = 
\rho_\Delta^S({\bf k}) \nonumber \\
& & \qquad
+ 2  e_\Delta a N^2 f_s(s) {\bf k}^2
\left[ \psi_S(\bar P,k) \psi_{D3}(\bar P,k) \right] 
\widehat
 Y_{20}(z) 
\nonumber \\
& & \qquad
 -  e_\Delta b^2 N^2
f_s(s)  {\bf k}^4|\psi_{D1}(\bar P,k)|^2 
\widehat
Y_{20}(z),
\label{eqRhoTotal}
\ea
where
$z=\cos \theta$, the function 
$\widehat Y_{20}(z)= \sfrac{1}{2}(3z^2-1)$ 
is proportional to the  spherical harmonic $Y_{20}$,
and   
\ba
\label{eq:rhokS}
& &
\rho_\Delta^S({\bf k})= e_\Delta N^2  \times \\
& &
\left[
|\psi_S(\bar P,k)|^2 +
a^2 {\bf k}^4 |\psi_{D3}(\bar P,k)|^2 
+ b^2 {\bf k}^4 |\psi_{D1}(\bar P,k)|^2
\right].
\nonumber
\ea

\begin{figure}[t]
\centerline{
\mbox{
\includegraphics[width=0.45\textwidth]{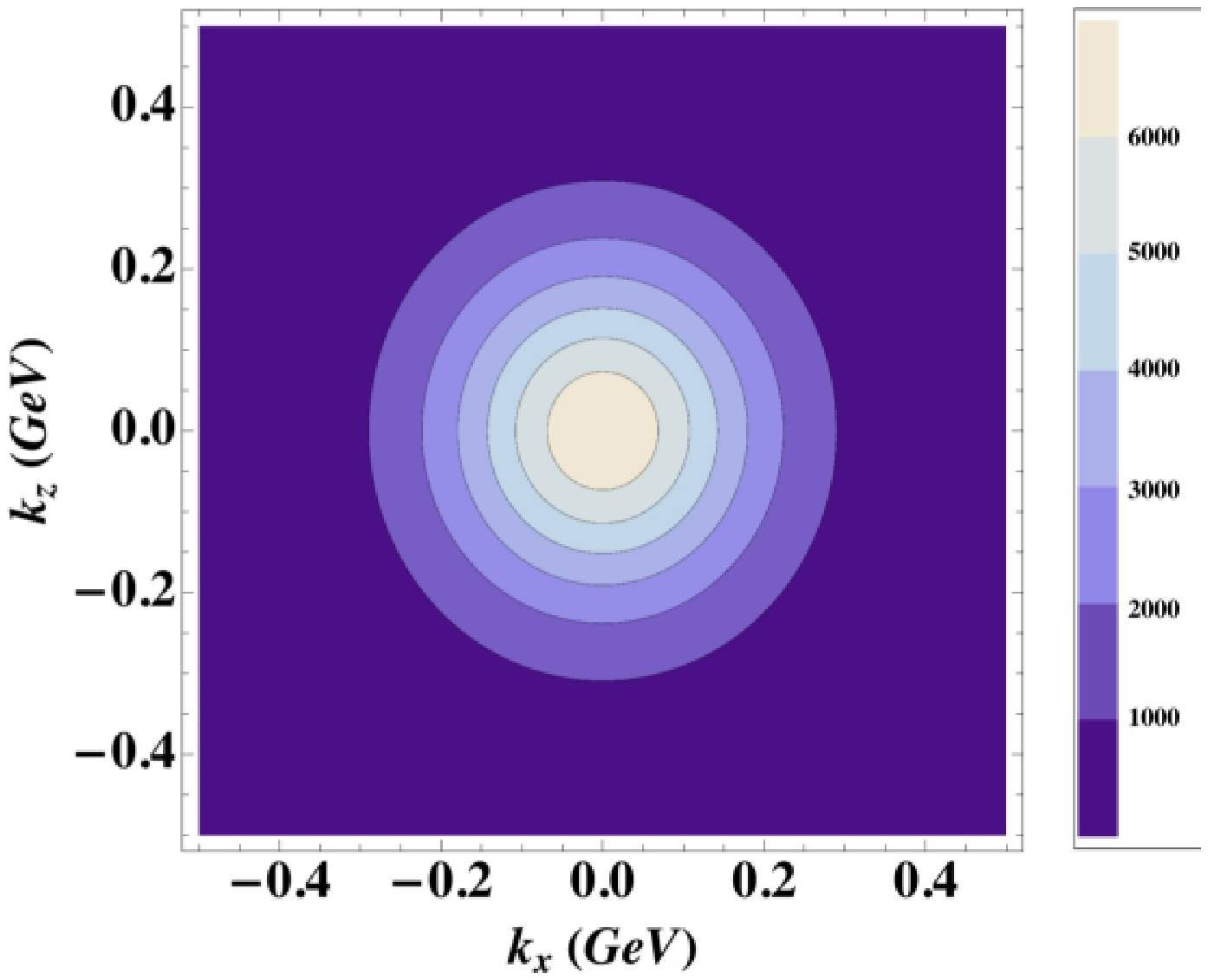}}}
\vspace{.3cm}
\centerline{
\mbox{
\includegraphics[width=0.45\textwidth]{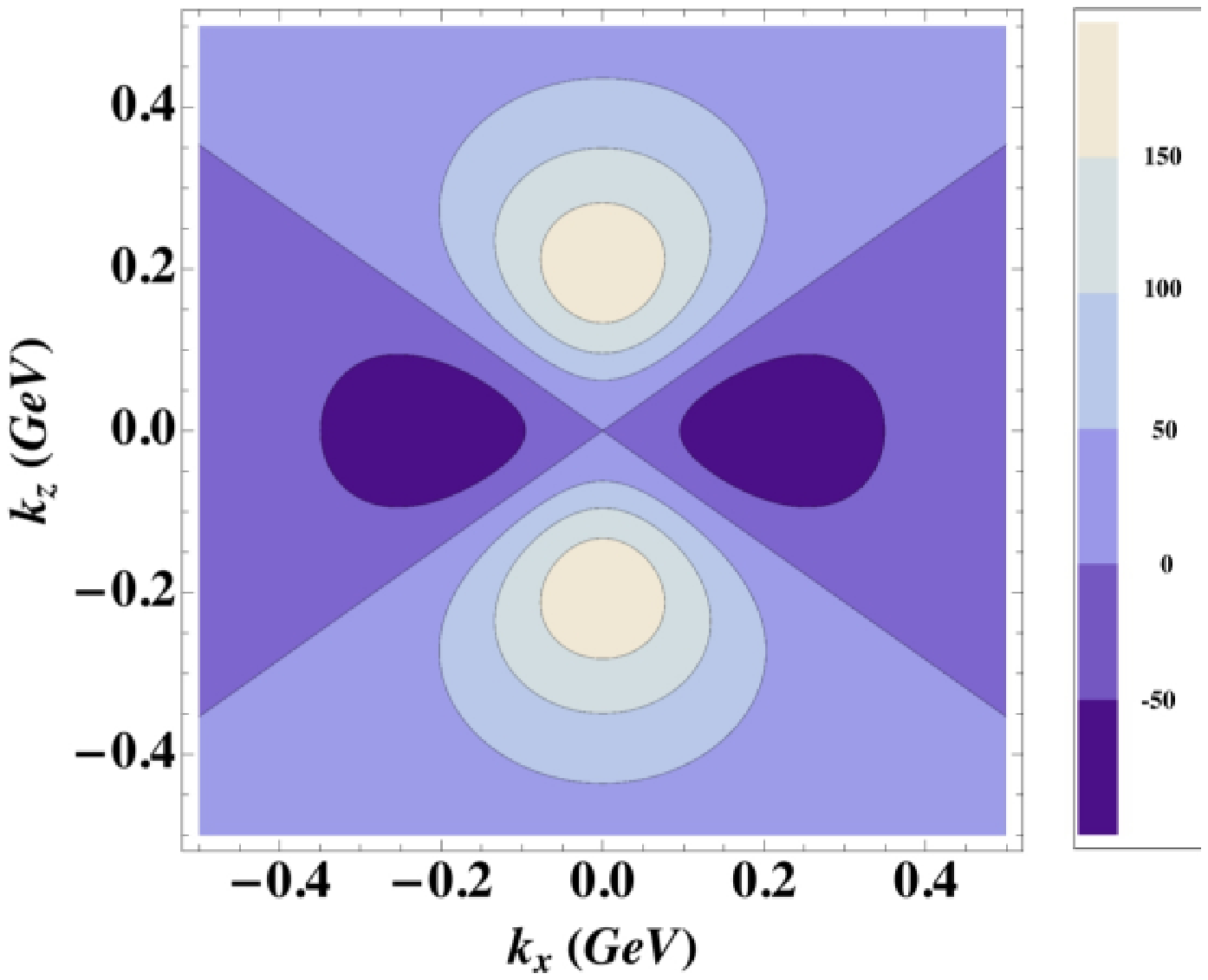}}}
\caption{\footnotesize{
Contour plots of momentum-space charge densities of the $\Delta$ 
in the $k_x$--$k_z$ plane, in units of GeV$^{-2}$. 
The top panel shows the total density 
$\rho_{\Delta}\left({\bf k},+\sfrac32\right)$. The bottom panel isolates the angle-dependent part 
$\rho^A_\Delta({\bf k} )$ induced by the $D$-states. Note the difference of the density scales in the two panels.
}}
\label{figK3D}
\end{figure}

\begin{figure}[t]
\centerline{
\mbox{
\includegraphics[width=0.45\textwidth]{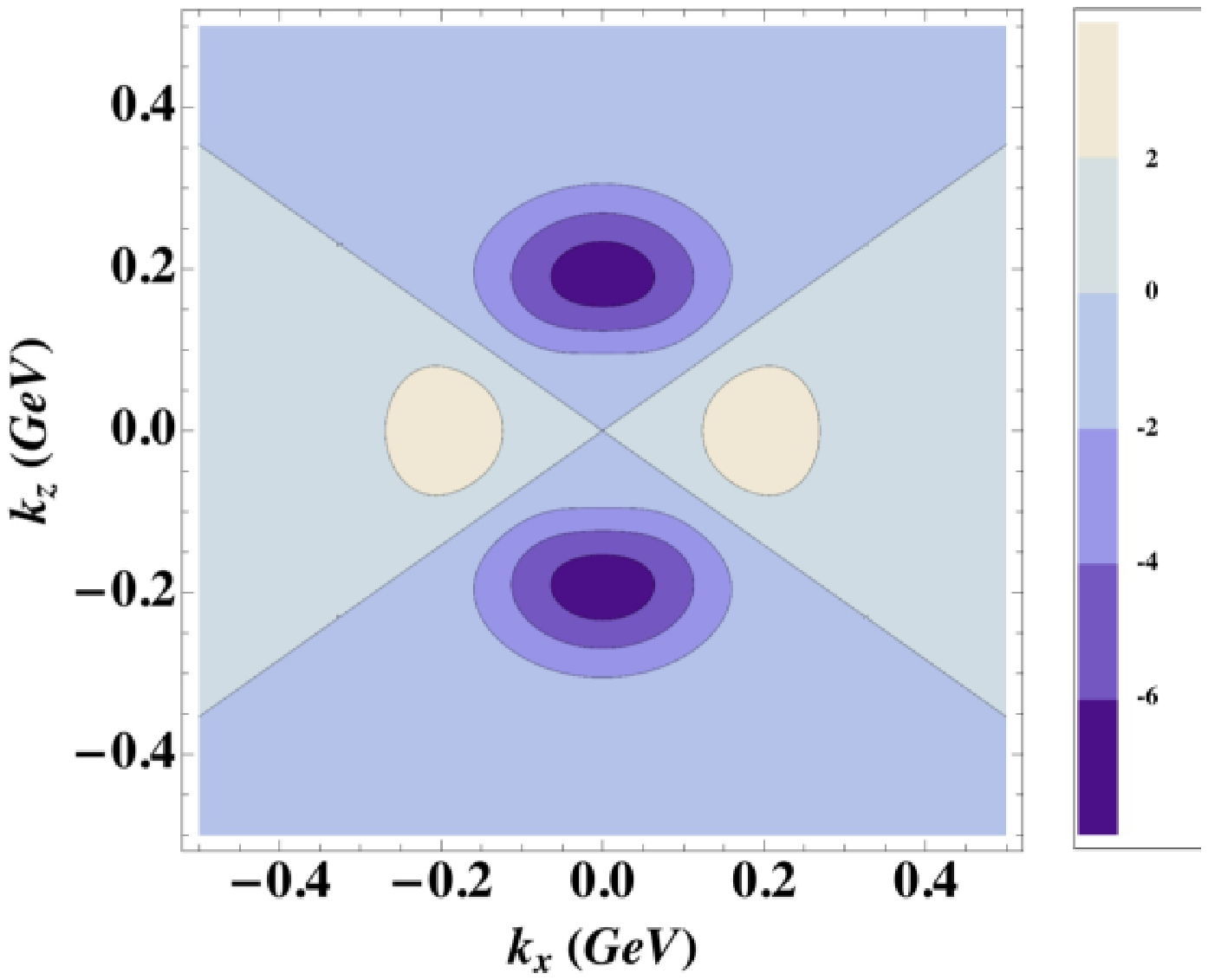}}}
\vspace{.3cm}
\centerline{
\mbox{
\includegraphics[width=0.45\textwidth]{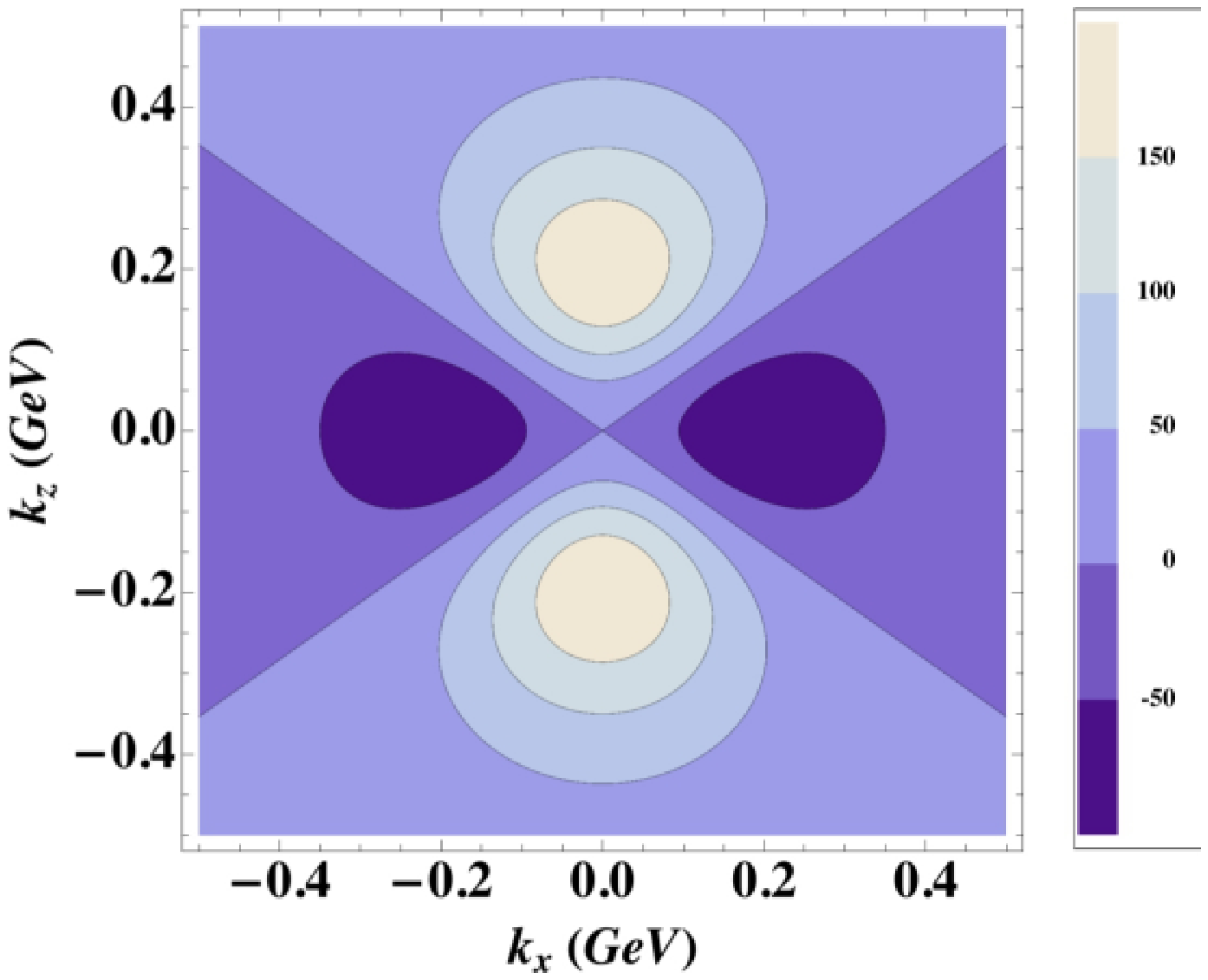}}}
\caption{\footnotesize{
Contour plots of momentum space charge densities of the $\Delta$ in the $k_x$--$k_z$ plane, in units of GeV$^{-2}$.
The upper panel shows the $D1-D1$ part, the lower panel the $S-D3$ part of  $\rho^A_\Delta ({\bf k} )$. Note the difference of the density scales in the two panels.
}}
\label{figK3D2}
\end{figure}

In the rest frame, $\psi_X$ depends only on ${\bf k}^2$.
The second term in (\ref{eqRhoTotal}) is the overlap 
between the $S$ and the $D3$ state. It does not vanish 
because the states have the same core spin-3/2. The $D1$ state on the other hand is always 
orthogonal to $S$ and $D3$ because its definition includes a 
spin-1/2 projector.
The last term comes from the overlap 
between the initial and final $D1$ states.
The two last terms 
of Eq.~(\ref{eqRhoTotal}) vanish when the  
angular integration is performed. 

The radially symmetric part of the density can also be 
written as 
\be
\rho_\Delta^S({\bf k})=
\sfrac{1}{2}
\left[ \rho_\Delta\left({\bf k},+\sfrac{3}{2}\right) + 
\rho_\Delta\left({\bf k},+\sfrac{1}{2}\right)
\right] \, , 
\label{eqRhoS}
\ee
and the angle-dependent asymmetric component as
\ba
\rho_\Delta^A({\bf k})&=&
\sfrac{1}{2}
\left[ \rho_\Delta({\bf k},+\sfrac{3}{2}) -
\rho_\Delta({\bf k},+\sfrac{1}{2})
\right] \nonumber \\
& = &
2 a e_\Delta N^2  {\bf k}^2 
\left[ \psi_S(\bar P,k) \psi_{D3}(\bar P,k) \right] \widehat Y_{20}(z)
\nonumber
\\
&&
 - b^2 e_\Delta N^2  {\bf k}^4|\psi_{D1}(\bar P,k)|^2 \widehat Y_{20}(z) .
\label{eqRhoA}
\ea
Using (\ref{eq:rhokS}) and (\ref{eqRhoA}) 
we can rewrite 
$\rho_\Delta({\bf k},s)$ as 
\ba
\rho_\Delta({\bf k},s)= 
\rho_\Delta^S({\bf k}) + f_s(s) \rho_\Delta^A({\bf k}). 
\label{eqNewRhoT}
\ea
From this equation and from 
$f_s\left(\pm \sfrac12\right)=-f_s\left(\pm \sfrac32\right)$ 
it is clear that the deformation density for 
$s=+ \sfrac12$  has always the opposite sign of the one for  $s=+ \sfrac32$.

For small $D$-state admixture coefficients $a$ and $b$, the 
$S$ to $D3$ transition term
dominates the asymmetry.
This is consistent with a 
calculation of the form factors 
in first order in $a$ and $b$ where the $D3$ state 
is responsible for a nonzero 
electric quadrupole form factor \cite{DeltaDFF,DeltaDFF2}.

The factors $\psi_S \psi_{D3}$ and $\psi_{D1}^2$  in Eq.~(\ref{eqRhoTotal}) are always greater than or equal to zero, and therefore the $D3$- and $D1$-state contributions to the deformation enter with opposite signs. The overall factor multiplying $\hat Y_{20}(z)$ can have either sign, depending on the specific parametrization of the wave functions and on the spin projection $s$. 

To illustrate the deformation of the $\Delta$ graphically, in Fig.~\ref{figKall}
we show a polar representation of $\rho_\Delta({\bf k},s)$, for 
${\bf k}$ in the $k_x$-$k_z$ plane (the densities are invariant under rotations about the $k_z$-axis). The positive $k_z$ direction, corresponding to polar angle $\theta=0$, points upwards. For fixed values of $\mbox{k}=|{\bf k}|$, the length of a straight line from the origin to a given point on a displayed curve is the respective density, and its angle with the upward direction is the polar angle $\theta$.
 
In this representation, a distribution $\rho_\Delta(\mbox{k},\theta; s)$ 
with no $\theta$-dependence yields  
a perfect circle. 
The three panels of Fig.~\ref{figKall} show  $\rho_\Delta({\bf k},s)$ for $\mbox{k}= 0.2$, $1.0$, and $2.0$ GeV, respectively. The perfect circle  (solid line) shows
the symmetric distribution $\rho_\Delta^S({\bf k})$.
In each case, the dashed line represents $s=+\sfrac32$, and the dotted line $s=+ \sfrac12$.
As determined by Eq.~(\ref{eqNewRhoT}) 
the deviations from a spherically symmetric density for 
$s=+ \sfrac12$ and $s=+ \sfrac32$ are equal but with opposite signs. 
In the case of the $s= + \sfrac32$ density, the deformation 
is along $k_z$ 
for the smaller momenta k=0.2 and k=1.0 GeV, but 
it is along $k_x$ 
for k=2.0 GeV.

This change in the shape of the deformation with increasing momentum
can be understood from the behavior of the $S-D3$ and $D1-D1$ 
terms in the density 
of Eq.~(\ref{eqRhoTotal}). Figure~\ref{figRho} shows the k-dependence 
of the magnitudes of $2a  N^2 {\bf k}^2 \psi_S \psi_{D3}$ and 
$b^2 {\bf k}^4 N^2 \psi_{D1}^2$, the factors that multiply $\hat{Y}_{20}$ 
with opposite signs, in comparison with 
the symmetric term $\rho_\Delta^S({\bf k})$. 
For small k, the $S-D3$ term dominates strongly over the $D1-D1$ term, which is expected already because it is of first order in the small $D$-wave admixture parameter $a$, whereas $D1-D1$ is quadratic in $b$. As k increases, the $S-D3$ contribution falls faster than $D1-D1$, due to the faster falloff of the $S$-state wave function.
Around $\mbox{k} \simeq 1.2$ GeV both contributions
become equal in magnitude and cancel against each other. For larger values of k $D1-D1$ dominates over $S-D3$, 
and changes the deformation from being along $k_x$ to being along $k_z$. 
However, the coefficients of $\hat Y_{20}$ at such high momenta 
are already very small (note the logarithmic scale in Fig.~\ref{figRho}).

Figure~\ref{figRho} shows also that, 
at high momenta, the spherically symmetric part of the density, $\rho_\Delta^S({\bf k})$, is itself dominated by its $D1-D1$ component. Comparing Eqs.\ (\ref{eqRhoTotal}) and  (\ref{eq:rhokS}) we see that the magnitude of the $D1-D1$ coefficient of $\hat Y_{20}(z)$ in (\ref{eqRhoTotal}), represented by the dotted line in Fig.~\ref{figRho}, is the same as the $D1-D1$ contribution to the spherically symmetric part of the density in (\ref{eq:rhokS}).  With increasing momentum k, the dotted line seems to converge to the solid line of the total symmetric contribution. However, it reaches only 
about 90\% of $\rho_\Delta^S({\bf k})$, the small $D3-D3$ contribution being responsible for the remainder. This ratio of the $D1-D1$ contribution to the total spherically symmetric part can be obtained using the asymptotic ratio 
$\left| \sfrac{\psi_{D3}}{\psi_{D1}} \right| 
\to \sfrac{1}{1-\lambda_{D1}} \sfrac{N_{D3}}{N_{D1}} \approx \sfrac13$.

Figure~\ref{figK3D} shows contour plots of  momentum space charge densities in the $k_x$--$k_z$ plane. 
The deformation of the total density 
$\rho_{\Delta}\left({\bf k},+\sfrac32\right)$, displayed in the upper panel, is barely visible in this plot, because the spherically symmetric contribution dominates strongly.
The lower panel shows only the much smaller asymmetric part 
$\rho^A_\Delta ({\bf k} )$, 
which enhances the density along the $k_z$-direction.

The origin of this enhancement is analyzed in Fig.~\ref{figK3D2}, 
where the upper panel shows the $D1-D1$, 
and the lower panel the $S-D3$ contribution. 
The $D1-D1$ channel density 
is deformed along $k_x$ 
but at the low momenta shown in the figure 
it is overwhelmed by the more than one order 
of magnitude larger $S-D3$ density deformation
along $k_z$.

In summary, at low and intermediate momenta the momentum space density 
distribution shows a small 
deformation in the $k_z$-direction
for the $\Delta$ state with spin projection $s=+\sfrac32$,
and a small deformation in the $k_z$-direction
for the $s=+ \sfrac12$ state. For high momenta the shape of the 
deformation is reversed, but this is hardly relevant 
because the density is already negligibly small in this region.

\subsection{Coordinate space}

We proceed now to the calculation of charge densities of the $\Delta$ in coordinate space. All densities will be determined in the rest frame of the $\Delta$, where the total four-momentum is $\bar P =(M_\Delta,{\bf 0})$.   Using the modified momentum-space wave function 
\be
\tilde \Psi_\Delta({\bf k}; s) \equiv
\frac{\Psi_\Delta(\bar P,k;s)}{\sqrt{2 E_s}} \, ,
\label{eqTilPsi}
\ee
where the relativistic phase-space factor is absorbed into the definition of the wave function,
we can write the  Fourier transform in the same form as for nonrelativistic wave functions:
\be
\tilde \Psi_\Delta({\bf r}; s) =
\int \frac{d^3 {\bf k}}{(2\pi)^3} e^{i {\bf k} \cdot {\bf r}}
\tilde \Psi_\Delta({\bf k}; s).
\label{eqFT}
\ee
With this convention for the factors of $2\pi$, the inverse transform is
\be
\tilde 
\Psi_\Delta({\bf k}; s) =
\int {d^3 {\bf r}} \; e^{ -i {\bf k} \cdot {\bf r} }
\tilde \Psi_\Delta({\bf r}; s).
\ee
In complete analogy to the momentum-space expression (\ref{eqRho}), the 
charge density in coordinate space is given by
\be
\tilde \rho_\Delta({\bf r};s)=
\sum_{\lambda_s} 
\tilde \Psi_\Delta^\dagger ({\bf r},s) j_q 
\tilde \Psi_\Delta ({\bf r},s).
\ee

As before, one can decompose $\tilde \Psi_\Delta({\bf r};s)$ 
into angular momentum components $S,D3,D1$.
In the fixed-axis polarization state basis we are using, there is no $k$-dependence in the diquark polarization vector 
$\left(\varepsilon_{\bar P}^\ast\right)_\alpha$. 
This makes the Fourier transform of the $S$-state (\ref{eqPsiS}) in the rest frame particularly simple, 
\be
\tilde
\Psi_S({\bf r};s)=
-  
R_S(\mbox{r}) 
\varepsilon_\alpha^\ast u^\alpha(\bar P,s),
\label{eqTilPsiSR}
\ee
where we have introduced the shorthand  $\varepsilon_\alpha^\ast $ to represent 
$\left(\varepsilon_{\bar P}^\ast\right)_\alpha$, and
\be
R_S(\mbox{r})
= \int \frac{d^3 {\bf k}}{(2\pi)^3}
e^{ i {\bf k} \cdot {\bf r} }
\frac{\psi_S(\bar P,k)}{\sqrt{2 E_s}},
\label{eqRS0}
\ee
with $\mbox{r}=|{\bf r}|$. The factor $\varepsilon_\alpha^\ast $
depends on the diquark polarization $\lambda_s$, which is not shown explicitly. The isospin states are not affected by the transformation and are also suppressed for simplicity.
There is no angle-dependence in the rest frame wave function $\psi_S(\bar P,k)$, therefore $R_S$ is also an $S$-wave, depending only on $\mbox{r}$. 
Boosting the wave function to another frame would induce 
an angle-dependence into $\psi_S(\bar P,k)$ and consequently also into $R_S$. 
However, this kind of relativistic deformation due to Lorentz contraction 
of the system along the direction of motion is a separate issue. 
We are interested in intrinsic deformations of the $\Delta$, 
which are already present in the rest frame.

Using the familiar expansion of a plane wave into partial waves, 
\ba
e^{i {\bf k} \cdot {\bf r}}=
4\pi \sum_{l=0}^{+ \infty}
\sum_{m=-l}^l i^l Y_{l m}(\hat {\bf r}) Y_{l m}^\ast (\hat {\bf k}) 
j_l(\mbox{k} \mbox{r}),
\label{eq:planewave}
\ea
where $j_l$ are the spherical Bessel functions, Eq.~(\ref{eqRS0}) becomes
\ba
R_S (\mbox{r})
= 4\pi 
\int \frac{\mbox{k}^2 d \mbox{k}}{(2\pi)^3}
j_0( \mbox{k} \mbox{r})
\frac{\psi_S(\bar P,k)}{\sqrt{2 E_s}}.
\label{eq:RS}
\ea

The general structure of the complete $D$-state wave functions (without isospin) can be written as 
\be
\tilde \Psi_D({\bf k};s) = 
\frac{\psi_D(\bar P,k)}{\sqrt{2 E_s}} \Phi_D(k,s),
\ee
where $D$ stands for $D1$ or $D3$.
$\Phi_D(k,s)$ is the spin wave function of the $D$-states.
In the rest frame, 
it can be represented as \cite{NDeltaD}
\ba
\Phi_{D1} (k,s) & = &
+\sqrt{4\pi} {\bf k}^2
\varepsilon_\alpha^\ast \nonumber \\
& & \times
\sum_{m_l s_1} 
\braket{2 m_l; \sfrac{1}{2} s_1}{\sfrac{3}{2} s} 
Y_{2 m_ l} (\hat {\bf k}) U^\alpha (\bar P,s_1) 
\nonumber \\
\Phi_{D3} (k,s) & = &
-\sqrt{4\pi} {\bf k}^2
 \varepsilon_\alpha^\ast \nonumber \\
&& \times
\sum_{m_l s_1}  
\left< 2 m_l; \sfrac{3}{2} s_1 | \sfrac{3}{2} s \right>
Y_{2 m_ l} (\hat {\bf k}) u^\alpha (\bar P,s_1)
\nonumber \\
\ea
where $U^\alpha$ is the spin-1/2 state \cite{NDelta,NDeltaD}
\ba
U^\alpha(P,s)=
\frac{1}{\sqrt{3}} \gamma_5 \left(
\gamma^\alpha - \frac{P^\alpha}{M}
\right) u(P,s).
\ea
The angle dependence is contained exclusively in the factors $Y_{2 m_ l}(\hat {\bf k})$, and the spin states are completely independent of ${\bf k}$.

The $D$-state wave functions in coordinate space
are then given by the Fourier transform, which yields
\ba
\tilde 
\Psi_{D1}({\bf r};s) &=&
+ \sqrt{4\pi}  \nonumber \\
&& \times 
\sum_{m_l s_1} 
\braket{2 m_l; \sfrac{1}{2} s_1}{\sfrac{3}{2} s} 
{\cal Y}_{D1}^{m_l} ({\bf r}) \left[ \varepsilon_\alpha^\ast 
U^\alpha(\bar P,s_1)
\right] \nonumber  \\
\tilde 
\Psi_{D3}({\bf r};s) &= &
-\sqrt{4\pi} \nonumber \\
& & \times
\sum_{m_l s_1} 
\braket{2 m_l; \sfrac{3}{2} s_1}{\sfrac{3}{2} s} 
{\cal Y}_{D3}^{m_l} ({\bf r}) \left[ \varepsilon_\alpha^\ast 
u^\alpha(\bar P,s_1)
\right], \nonumber \\
& &
\label{eq:PsiD13}
\ea
where
\be
{\cal Y}_{D}^{m_l} ({\bf r}) 
=\int \frac{d^3 {\bf k} }{(2\pi)^3} 
e^{ i {\bf k} \cdot {\bf r} }{\bf k}^2 Y_{2 m_l}(\hat {\bf k}) 
\frac{\psi_{D}(\bar P,k)}{\sqrt{2 E_s}}.
\ee
Again, in the rest frame ${\psi_{D}(\bar P,k)}$ is also 
angle independent, and the Fourier transform simplifies to
\ba
{\cal Y}_{D}^{m_l} ({\bf r}) =
-R_D(\mbox{r}) Y_{l m_l} (\hat {\bf r}),
\label{eq:YD}
\ea
where
\ba
R_D (\mbox{r})=  4\pi 
\int \frac{\mbox{k}^2 d \mbox{k}}{(2\pi)^3} 
\mbox{k}^2 j_2( \mbox{k} \mbox{r}) \frac{\psi_D(\bar P,k)}{\sqrt{2 E_s}}, 
\label{eqR}
\ea
and the minus sign factored out in (\ref{eq:YD}) comes from the $i^l$ 
in (\ref{eq:planewave}).

\begin{figure}[t]
\centerline{
\mbox{
\includegraphics[width=3.3in]{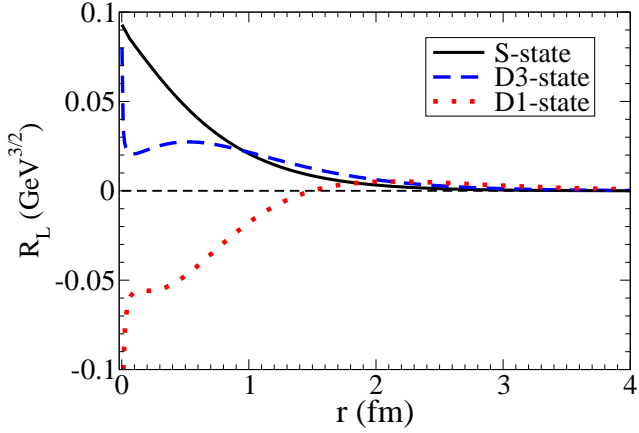}}}
\caption{\footnotesize{Radial $S$- and $D$-state wave functions of the $\Delta$ in coordinate space, calculated through Fourier transforms according to Eqs.~(\ref{eq:RS}) and (\ref{eqR}). }}
\label{figFourier1}
\end{figure}

We calculated the functions $R_S$, $R_{D3}$, and 
$R_{D1}$ numerically, and the results 
are presented in Fig.~\ref{figFourier1}.
The $S$-state dominates at small distances, 
but the two $D$-state wave functions become 
comparable in size around $\mbox{r} \approx 1$ fm 
and dominate for larger values of $\mbox{r} $. 
The $D$-waves start out with opposite signs, but the $D1$ 
wave 
changes sign at $\mbox{r} \approx 1.4$ fm.

As r goes to zero, the relativistic $D$-state wave functions 
are weakly singular, 
namely $R_D(\mbox{r}) \propto \mbox{r}^{-1/2}$. 
This behavior, which is reminiscent of the singular radial dependence 
of the Dirac wave functions 
of the hydrogen atom \cite{BetheSalpeter}, 
does not cause any problems because the densities remain integrable. 
The origin of these singularities is the slower falloff 
with increasing relative momentum 
of the relativistic momentum-space wave functions 
compared to nonrelativistic wave functions.
If we calculate 
the Fourier transform of the $D$-state wave functions 
in the nonrelativistic limit, at small r we obtain 
the regular behavior $R_D(\mbox{r})  \propto \mbox{r}^2$.

\begin{figure*}[t]
\centerline{
\mbox{
\includegraphics[width=2.0in]{Fig6a} \hspace{.3cm}
\includegraphics[width=2.1in]{Fig6b} \hspace{.3cm}
\includegraphics[width=2.0in]{Fig6c}}}
\caption{\footnotesize{Polar plots of $\tilde \rho_\Delta ({\bf r})$ 
for three fixed values of r $=|{\bf r}|$.
In each case, the solid line represents $\tilde \rho_{\Delta}^S({\bf r})$,
the dashed line $\tilde \rho_\Delta\left({\bf r},+\sfrac32\right)$,  
and the dotted line $\tilde \rho_\Delta\left({\bf r},+\sfrac12 \right)$.
The scale for $\rho_\Delta\left({\bf r},s\right)$ along 
the $x$ and $z$ axes is in units of GeV$^{3}$.  
}}
\label{figRall}
\end{figure*}

\begin{figure}[t]
\vspace{.3cm}
\centerline{
\mbox{
\includegraphics[width=3.3in]{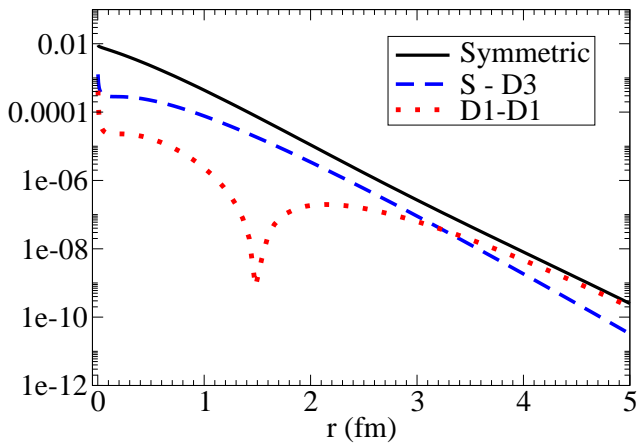}}}
\caption{\footnotesize{
Comparison of the three contributions to the total coordinate-space density $\tilde \rho_\Delta({\bf r},s)$  in Eq.~(\ref{eqRho3}) in units of GeV$^3$.
The solid line represents the symmetric contribution, 
$\tilde\rho^S_\Delta ({\bf r})$, the dashed and dotted lines show 
the coefficients
of $\hat{Y}_{20}(z)$ proportional to $R_S R_{D3}$ and $R_{D1}^2$, 
respectively. 
In all cases, the common factor $e_\Delta=1$,
and only the absolute values are plotted.
}}
\label{figRhoL}
\end{figure}

The total coordinate-space
charge density is
\ba
& &
\tilde \rho_\Delta({\bf r},s) =
N^2 \tilde \rho_{\Delta, S}({\bf r},s) 
\nonumber \\
& & \qquad +
a^2 N^2 \tilde \rho_{\Delta ,D3}({\bf r},s)+ 
b^2 N^2 \tilde \rho_{\Delta ,D1}({\bf r},s)  \nonumber \\
& & \qquad + 2 aN^2 \tilde \rho_{\Delta ,SD3} ({\bf k},s).
\label{eqRho2}
\ea
The various components are 
defined in analogy with (\ref{eqRhoX}) and (\ref{eqRhoSD3}).

The density associated with the $S$-state is
\ba
\tilde \rho_{\Delta, S}({\bf r},s)= e_\Delta 
R_S^2 \, .
\ea
For the other cases, one uses
\ba
& &
\sum_{\lambda_s} 
\varepsilon_{P}^\alpha (\lambda_s) 
\varepsilon_{P}^{\beta \ast} (\lambda_s)=
- g^{\alpha \beta} +  \frac{P^\alpha P^\beta}{M_\Delta^2}  \, ,\\ 
& &
\bar u_\alpha(P,s) u^\alpha(P,s) = 
\bar U_\alpha (P,s)U^\alpha (P,s)= -1 \, , \qquad \\
& &
\bar u_\alpha (P,s) U^\alpha(P,s)= \bar U^\alpha(P,s) u_\alpha(P,s) =0\, , 
\qquad  
\ea
assuming the same polarization in the initial
and final states.
The results for $D3$, $D1$, 
and the transition $S$ to $D3$ are
\ba
& &
\tilde \rho_{\Delta, D3} \Big( {\bf r}, +\sfrac{3}{2} \Big)= 
e_\Delta 
R_{D3}^2  \, ,\\
& &
\tilde \rho_{\Delta, D1}\Big( {\bf r}, +\sfrac{3}{2}\Big)= 
e_\Delta R_{D1}^2  
\left[1 - \widehat Y_{2\, 0}(z) \right] \, ,\\
& &
\tilde \rho_{\Delta, S\, D3}\Big( {\bf r} , +\sfrac{3}{2}\Big)= 
- 
e_\Delta R_{D3}
R_S \widehat Y_{2\,0}(z),
\ea
for $s=+ \sfrac32$, and
\ba
& &
\tilde \rho_{\Delta, D3}\Big( {\bf r} , +\sfrac{1}{2} \Big)= 
e_\Delta 
R_{D3}^2  \, ,\\
& &
\tilde \rho_{\Delta, D1}\Big( {\bf r} , +\sfrac{1}{2}\Big)= 
e_\Delta R_{D1}^2 
\left[1 +  \widehat Y_{2\,0}(z)\right] \, ,\\
& &
\tilde \rho_{\Delta, S\, D3}\Big( {\bf r} , +\sfrac{1}{2}\Big)= 
e_\Delta R_{D3}
R_S \widehat Y_{2\,0} (z) \, ,
\ea
for $s=+ \sfrac12$. The function 
$\widehat Y_{2\,0}(z)$, with $z=\cos \theta$, was defined previously, but 
the angle $\theta$ is now to be understood as the angle 
between $\hat {\bf r}$ and the $z$-axis.
More details are given in Appendix \ref{apRho}.

The total density becomes
\be
\tilde \rho_\Delta({\bf r},s) = \tilde \rho_{\Delta}^S({\bf r}) 
+ f_s(s) \tilde \rho_{\Delta}^A({\bf r}) ,
\ee
where
\ba
\hspace{-.5cm}
\tilde \rho_{\Delta}^S({\bf r}) &=&
N^2  e_\Delta 
\left[ R_S^2 +
a^2 R_{D3}^2 + b^2 R_{D1}^2 \right]
\ea
is again the angle- and spin-projection-independent contribution, and
\ba
\tilde \rho_{\Delta}^A({\bf r})&=&
 - 2  e_\Delta
a N^2   R_S R_{D3}
\widehat Y_{2\, 0} (z) \nonumber \\
& &
-  e_\Delta
b^2 N^2   R_{D1}^2  \widehat Y_{2\, 0} (z),
\label{eqRho3}
\ea
the angle-dependent asymmetric component. Equation~(\ref{eqRho3}) is the coordinate-space 
analogue of Eq.~(\ref{eqRhoA}).
The different sign of the term containing the $S$ and $D3$ wave functions 
is due to Eq.~(\ref{eq:YD}) where $R_D(\mbox{r})$ 
is defined such that it does not contain the factor $i^l$.

\begin{figure}[t]
\vspace{.3cm}
\centerline{
\mbox{
\includegraphics[width=0.45\textwidth]{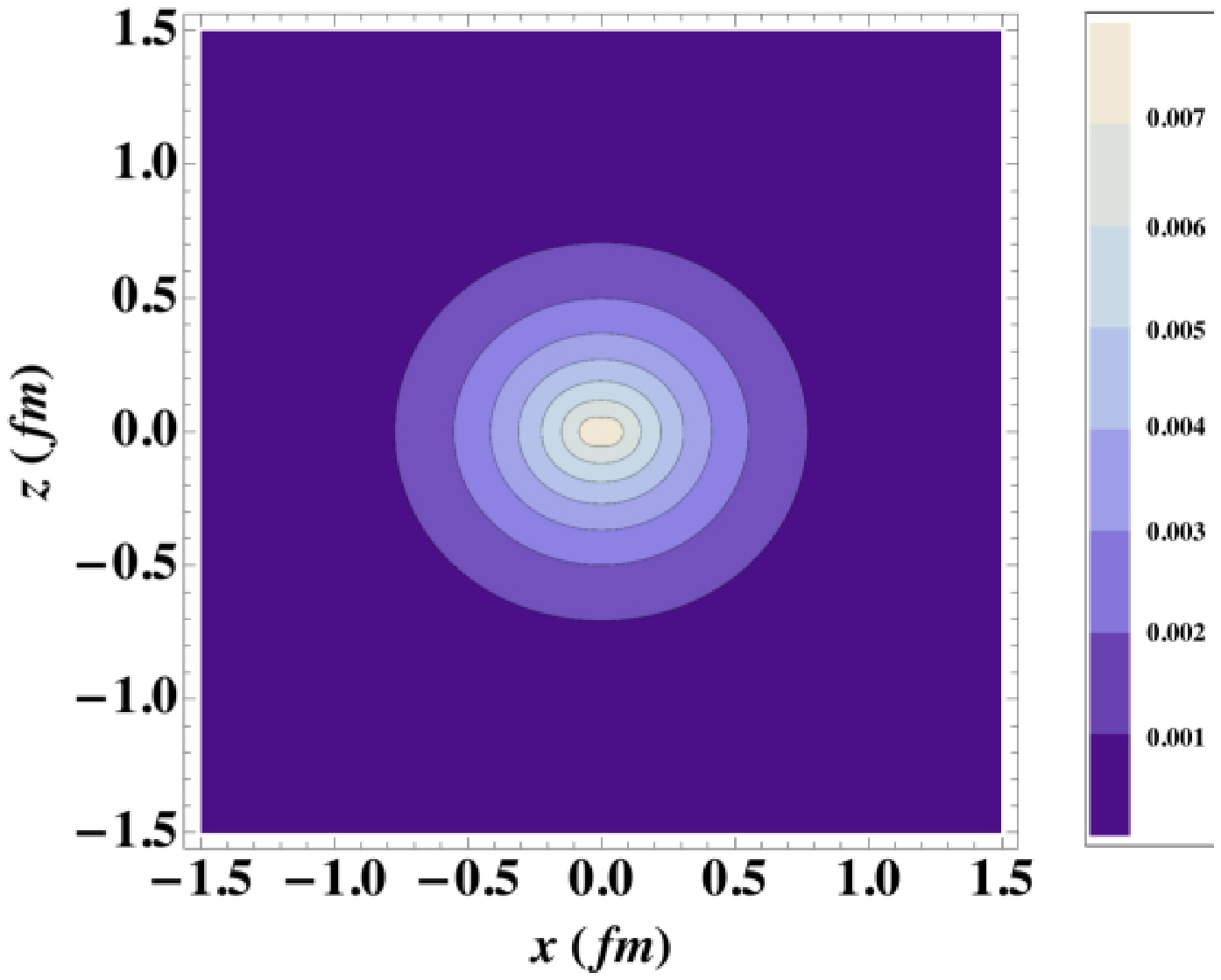}}}
\centerline{
\mbox{
\includegraphics[width=0.45\textwidth]{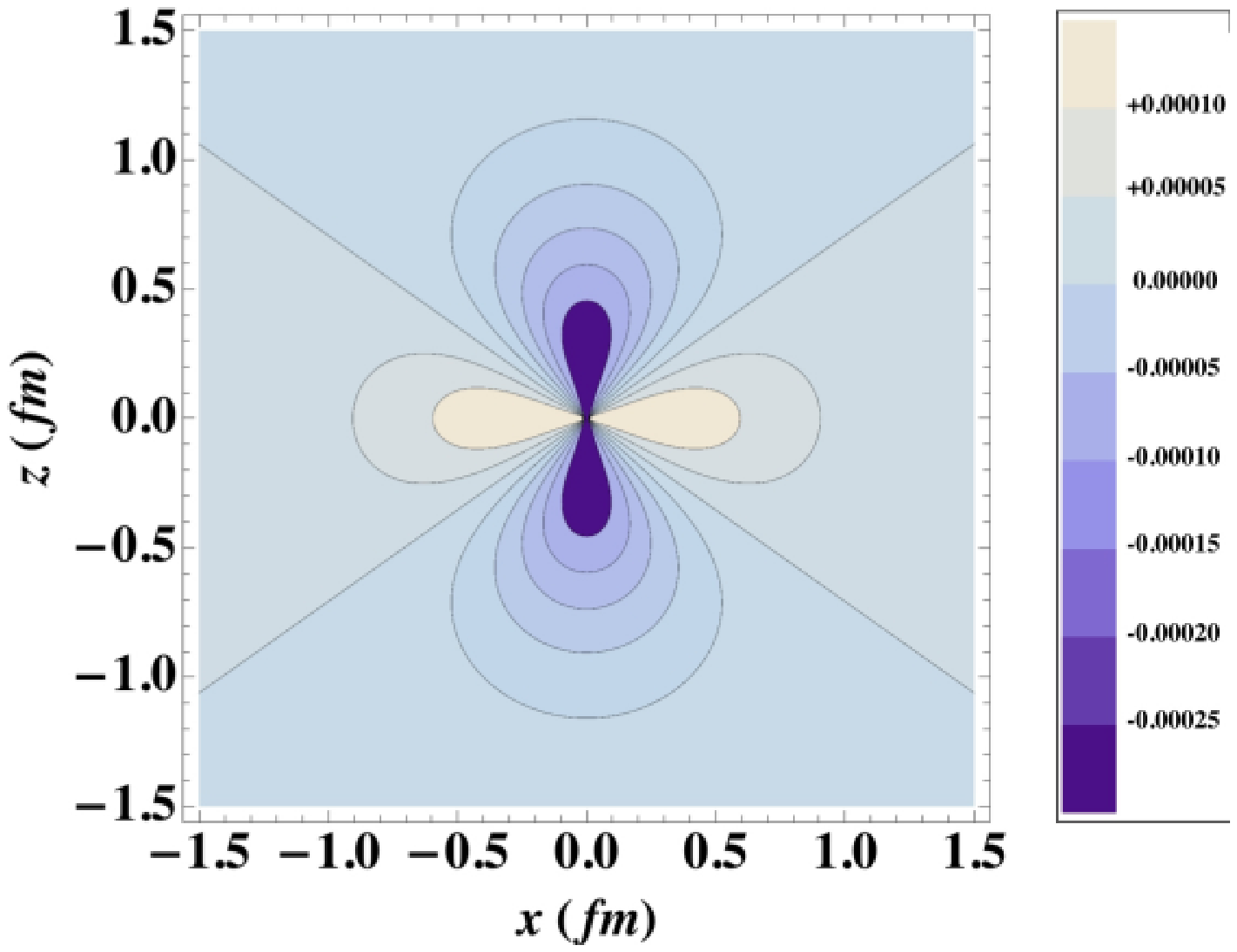}}}
\caption{\footnotesize{
Contour plots of coordinate-space charge densities of the $\Delta$ 
in the $x$--$z$ plane in units of GeV$^{3}$. 
The top panel shows the total density 
$\tilde \rho_{\Delta}\left({\bf r},+\sfrac32\right)$. 
The bottom panel isolates the angle-dependent part 
$\tilde\rho^A_\Delta ({\bf r} )$ induced by the $D$-states.
}}
\label{figSpace3D}
\end{figure} 

Again there are two independent terms proportional 
to $\widehat Y_{2\, 0}(z)$ that cause 
deformation, one associated with the $D1$ state, and another 
with a $S$ to $D3$ transition.
The main difference to the analogous momentum-space expression 
is that both terms come now with a minus sign.
Because $R_{D1}$ enters squared, the total effect of the 
two $D$-state contributions can be inferred 
from the sign of the wave functions $R_S$ and $R_{D3}$
and the sign of $a$.

When $a$ is positive, as in model Spectator-SD,
and  $R_S$, $R_{D3}$ have the same sign, 
the two terms in Eq.~(\ref{eqRho3}) have also 
the same sign and reinforce each other's contribution to the deformation.
This is the case 
in the region up to $\mbox{r} =  4$ fm shown in Fig.~\ref{figFourier1}, 
where both $R_S$ and $R_{D3}$ are positive.
We checked that both $R_S$ and $R_{D3}$ stay positive in that region when the 
wave function parameters are varied within a broad range, such that the direction of the deformation 
remains a robust result.

For  $s=+ \sfrac32$ this means that $\widehat Y_{20}(z)$ 
is multiplied by an overall negative factor,
implying an oblate shape, as shown in 
Fig.~\ref{figRall} for r = 0.2, 1.0, and 2.0 fm.  
For $s=+\sfrac12$ the coefficient of $\widehat Y_{20}(z)$ 
has the opposite sign, and the deformation is prolate.
At  $\mbox{r}=0.2$ fm the deformation is too small to be visible in 
the graph, but it becomes more pronounced as r increases.

Figure \ref{figRhoL} shows the magnitudes of the individual coefficients of $\widehat Y_{20}(z)$, together with the radially symmetric $\tilde \rho_{\Delta}^S({\bf r})$. The $S-D3$ contribution dominates
over  $D1-D1$ at small r, because it is of first order in the small 
$D$-state admixture coefficient $a$, whereas $D1-D1$ is of 
second order in $b$. 
However, with increasing r the $S$-wave falls more rapidly than 
the $D$-waves, and the $D1$-term dominates for $\mbox{r} >3$ fm.
  
Finally, Fig.~\ref{figSpace3D} presents contour plots 
of the charge densities of the $s=+\sfrac32$ state in the $x$-$z$ plane.
The upper panel shows the total density, 
the lower panel only the much smaller asymmetric contribution 
which is responsible for the oblate shape.

Our results for the shape  of the $\Delta^+$ 
are in  agreement with those of
previous studies,  such as the results reported in Ref.~\cite{Gross83} 
using the cloudy bag model, and the constituent quark model calculations of Ref.~\cite{Buchmann00}.
The oblate shape for $\Delta^+$ is also consistent  
with the negative sign of $G_{E2}(0)$
found in lattice QCD simulations
\cite{Alexandrou09,Boinepalli09}.

The formalism presented here can be applied to other systems. 
For instance, from an analysis of the form factors of the  $\Omega^-$  baryon, it can be concluded that the charge density distribution of the $\Omega^-$ also has an oblate shape  \cite{GE2Omega}.

\section{Summary and conclusions}
\label{secConclusions}

The $\Delta$ is the lowest-mass baryon that can possess a nonvanishing electric quadrupole moment.
In a nonrelativistic framework, the electric quadrupole moment can be used as an indicator for a particle's deviation from a spherically symmetric shape. However, in the general, relativistic case the connection between shape and higher spectroscopic moments is more complicated, which led to the proposal of alternative methods to measure deformation. 

One of these methods suggests to extract information about 
deformation from transverse densities, 
calculated in the transverse impact-parameter space $(b_x,b_y)$ 
in the infinite-momentum frame. 
It has the advantage 
that the transverse density moments 
${\cal Q}_\Delta^\perp$ and ${\cal O}_\Delta^\perp$ 
are zero for a structureless particle in the transverse plane,
and therefore can be used to measure the extension 
of the particle in impact-parameter space.
However, they do not allow the classification of 
the shape in the particle's rest frame.

In this work we used a different relativistic formalism, 
the covariant spectator theory, to investigate 
the relation between moments of charge 
or magnetic density distributions and 
the intrinsic shape of these distributions in the 
$\Delta$ baryon's rest frame.  
We used two covariant quark-diquark momentum-space wave functions 
for the $\Delta$, one consisting of pure $S$-waves only, called 
``Spectator-S'', and another which includes $S$- and $D$-waves, 
called ``Spectator-SD''. The electric and magnetic moments 
and form factors can be calculated directly from these wave functions, 
as well as the momentum space densities. 
The coordinate space densities are then obtained from the 
Fourier-transformed wave functions. 

We arrived at the following results:

For the $S$-wave model Spectator-S one obtains $G_{E2}(0)=G_{M3}(0) = 0$, 
and the electric and magnetic density distributions are spherically symmetric.

The $D$-wave admixture in model Spectator-SD on the other hand produces 
a spatial deformation of the 
$\Delta^+$ density distribution,
and the quadrupole and octupole moments become 
$G_{E2}(0) = -1.70$ and $G_{M3}(0) = -1.72$, respectively. 
The negative value of the quadrupole moment corresponds to 
an oblate density distribution in coordinate space.
We conclude therefore that the higher moments are good indicators that allow to distinguish deformed from spherically symmetric systems. 

Using the same wave functions and their respective electric and magnetic moments, we also calculated the corresponding transverse density quadrupole and octupole moments (${\cal Q}_\Delta^\perp$ and ${\cal O}_\Delta^\perp$). For each moment, the obtained values (see Table \ref{tableQdel}) for models Spectator-S and 
Spectator-SD are nonvanishing and of the same sign, and thus do not show a clear distinction between spherically symmetric and deformed cases. 
Since the transverse moments are nonzero one can conclude that the 
system is not seen as a point 
in impact parameter 
space when viewed from the light-front frame, 
but it is not clear how further information 
on its shape could be extracted.
In this sense, the usual three-dimensional density distribution complements the information contained in the transverse densities, and it is also closer to our intuitive notion of deformation.

For the specific case of the $\Delta^+$ baryon with spin projection 
$s= + \sfrac32$, 
the covariant model Spectator-SD 
predicts an oblate shape of its density distribution in coordinate space. 
This is in agreement with previously obtained results, 
both from other quark model calculations and lattice QCD simulations.

All model parameters were determined through fits to the available 
lattice QCD data for the $\gamma N \rightarrow \Delta$ 
transition form factors at large pion mass, 
where the uncertain pion cloud effects are minimal 
(the experimental data are also well predicted)
\cite{LatticeD}. 
In particular, the coefficients $a$ and $b$ are determined 
by the lattice transition form factors for $Q^2 \approx 0$, namely 
$a$ is determined by the 
electric quadrupole form factor $G_E^\ast(Q^2)$
and  $b$ by the Coulomb quadrupole form factor $G_C^\ast(Q^2)$.
Further improvements in the statistical quality of the lattice data might alter the magnitudes of $a$ and $b$ and therefore the extent of the deformation we predict for the $\Delta$, but the signs are not in doubt. 
In this sense, the nature of the deformation, 
namely that $\Delta^+$ 
is oblate rather than prolate, 
is a robust prediction of our model.

\acknowledgments

G.\ R.\ thanks Carl Carlson for helpful discussions.
This work was supported in part by the European Union under the 
HadronPhysics3 Grant No.~283286, 
and by the Funda\c{c}\~ao para a Ci\^encia e a
Tecnologia (FCT), under Grant No.~PTDC/FIS/113940/2009. One of us (G.\ R.) was also supported
by FCT under Grant No.~SFRH/BPD/26886/2006.

\appendix

\vspace{.2cm}

\section{Explicit expressions for 
$\tilde \Psi_{D1} ({\bf r}; s)$ and 
 $\tilde \Psi_{D3} ({\bf r}; s)$}
\label{apRho}

We list here more explicit expressions of the coordinate-space $\Delta$ wave functions $\tilde \Psi_{D1} ({\bf r}; s)$ and 
$\tilde \Psi_{D3} ({\bf r}; s)$. Performing the sums in Eq.\ (\ref{eq:PsiD13}) we get

\ba
\tilde \Psi_{D1} \left({\bf r}; +\sfrac{3}{2} \right)&=&
- \sqrt{\frac{4\pi}{5}}  {\cal Y}_{D1}^{+1} ({\bf r}) 
\varepsilon_\alpha^\ast \,
U^\alpha\left( \bar P, + \sfrac{1}{2}\right) 
\nonumber  \\
& &
+ \sqrt{\frac{16\pi}{5}}  {\cal Y}_{D1}^{+2} ({\bf r}) 
\varepsilon_\alpha^\ast \,
U^\alpha\left(\bar P, - \sfrac{1}{2}\right), 
\\
& & \nonumber \\
\tilde \Psi_{D1} \left({\bf r}; +\sfrac{1}{2} \right)&=&
- \sqrt{\frac{8\pi}{5}}  {\cal Y}_{D1}^{0} ({\bf r}) 
\varepsilon_\alpha^\ast \,
U^\alpha\left( \bar P,+ \sfrac{1}{2}\right) 
\nonumber  \\
& &
+ \sqrt{\frac{12\pi}{5}}  {\cal Y}_{D1}^{+1} ({\bf r}) 
\varepsilon_\alpha^\ast \,
U^\alpha\left( \bar P,- \sfrac{1}{2}\right).
\ea

and
\ba
\tilde \Psi_{D3} \left({\bf r}; +\sfrac{3}{2} \right)&=&
- \sqrt{\frac{4\pi}{5}}  {\cal Y}_{D3}^0 ({\bf r}) \varepsilon_\alpha^\ast 
u^\alpha\left( \bar P,+ \sfrac{3}{2}\right) 
\nonumber  \\
& &
+ \sqrt{\frac{8\pi}{5}}  {\cal Y}_{D3}^{+1} ({\bf r}) \varepsilon_\alpha^\ast 
u^\alpha\left( \bar P,+ \sfrac{1}{2}\right)  \nonumber \\
& &
- \sqrt{\frac{8\pi}{5}}  {\cal Y}_{D3}^{+2} ({\bf r}) \varepsilon_\alpha^\ast 
u^\alpha\left( \bar P,- \sfrac{1}{2}\right),  
\\
& & \nonumber \\
\tilde \Psi_{D3} \left({\bf r}; +\sfrac{1}{2} \right)&=&
- \sqrt{\frac{8\pi}{5}}  {\cal Y}_{D3}^{-1} ({\bf r}) \varepsilon_\alpha^\ast 
u^\alpha\left( \bar P,+ \sfrac{3}{2}\right) 
\nonumber  \\
& &
+ \sqrt{\frac{4\pi}{5}}  {\cal Y}_{D3}^{0} ({\bf r}) \varepsilon_\alpha^\ast 
u^\alpha\left( \bar P, + \sfrac{1}{2}\right)  \nonumber \\
& &
- \sqrt{\frac{8\pi}{5}}  {\cal Y}_{D3}^{+2} ({\bf r}) \varepsilon_\alpha^\ast 
u^\alpha\left( \bar P,- \sfrac{3}{2}\right).  
\ea

In the calculation of the density distributions from the coordinate wave functions  
we used the following relations:
\ba
& &
4 \pi 
\left\{\frac{2}{5} 
\left|Y_{2\,+2}(\hat {\bf r})\right|^2 +
\frac{2}{5} 
\left|Y_{2\,+1}(\hat {\bf r})\right|^2 +
\frac{1}{5} 
\left| Y_{2\,0}(\hat {\bf r})\right|^2 
\right\} = 1 \nonumber \\
& &
4 \pi 
\left\{\frac{2}{5} 
\left|Y_{2\,+2}(\hat {\bf r})\right|^2 +
\frac{1}{5} 
\left|
Y_{2\,0}(\hat {\bf r})\right|^2 +
\frac{2}{5} 
\left|
Y_{2\,-1}(\hat {\bf r})\right|^2 
\right\}=1, \nonumber \\
& & \\
& &
4 \pi 
\left\{\frac{4}{5} 
\left|Y_{2\,+2}(\hat {\bf r})\right|^2 +
\frac{1}{5} 
\left|Y_{2\,+1}(\hat {\bf r})\right|^2
\right\} 
= 1 - \frac{1}{2}\left( 3 z^2 -1 \right)
\nonumber \\
& &
4 \pi 
\left\{\frac{3}{5} 
\left|Y_{2\,+1}(\hat {\bf r})\right|^2 +
\frac{2}{5} 
\left|
Y_{2\,0}(\hat {\bf r})\right|^2
\right\}
= 1 +  \frac{1}{2}\left( 3 z^2 -1 \right). \nonumber \\
& &
\ea

\end{document}